\documentclass[aps,prd,twocolumn,nofootinbib,superscriptaddress,floatfix]{revtex4-1}
\usepackage{graphicx}
\usepackage{bm}
\usepackage{amsmath}
\usepackage{amssymb}
\usepackage{amsfonts}
\usepackage{hyperref}
\usepackage{bbold}
\hypersetup{colorlinks=true,citecolor=blue,linkcolor=blue,urlcolor=blue}
\usepackage{tabularx}
\usepackage{xspace}
\usepackage[normalem]{ulem}
\usepackage{soul}
\usepackage[dvipsnames]{xcolor}

\def\pasp{\ref@jnl{PASP}} 
\def\physrep{\ref@jnl{Phys.~Rep.}}   

\def\apj{ApJ}

\def\physrep{Phys. Rep.}

\def\pasp{PASP}
\newcommand*{\Euclid}{\textit{Euclid}\xspace}

\begin{document}

\title{Constraining neutrino masses with weak-lensing multiscale peak counts}

\author{Virginia Ajani}\email{virginia.ajani@cea.fr}
\affiliation{AIM, CEA, CNRS, Universit{\'e} Paris-Saclay, Universit{\'e} Paris Diderot, 
             Sorbonne Paris Cit{\'e}, F-91191 Gif-sur-Yvette, France}
\author{Austin Peel}
\affiliation{Laboratoire d'Astrophysique, Ecole Polytechnique F{\'e}d{\'e}rale de Lausanne (EPFL),
             Observatoire de Sauverny, CH-1290 Versoix, Switzerland}

\author{Valeria Pettorino}
\affiliation{AIM, CEA, CNRS, Universit{\'e} Paris-Saclay, Universit{\'e} Paris Diderot, 
             Sorbonne Paris Cit{\'e}, F-91191 Gif-sur-Yvette, France}
\author{Jean-Luc Starck}
\affiliation{AIM, CEA, CNRS, Universit{\'e} Paris-Saclay, Universit{\'e} Paris Diderot, 
             Sorbonne Paris Cit{\'e}, F-91191 Gif-sur-Yvette, France}
\author{Zack Li}
\affiliation{Department of Astrophysical Sciences, Princeton University, 4 Ivy Lane, Princeton, NJ 08544, USA}

\author{Jia Liu}
\affiliation{Berkeley Center for Cosmological Physics, University of California, Berkeley, CA 94720, USA}
\affiliation{Department of Astrophysical Sciences, Princeton University, 4 Ivy Lane, Princeton, NJ 08544, USA}

\begin{abstract}
\noindent Massive neutrinos influence the background evolution of the Universe as well as the growth of structure. Being able to model this effect and constrain the sum of their masses is one of the key challenges in modern cosmology.  Weak lensing cosmological constraints will also soon reach higher levels of precision with next-generation surveys like LSST, WFIRST and \Euclid. In this context, we use the \url{MassiveNus} simulations to derive constraints on the sum of neutrino masses $M_{\nu}$, the present-day total matter density $\Omega_{\rm m}$, and the primordial power spectrum normalization $A_{\rm s}$ in a tomographic setting. We measure the lensing power spectrum as second-order statistics along with peak counts as higher-order statistics on lensing convergence maps generated from the simulations. We investigate the impact of multi-scale filtering approaches on cosmological parameters by employing a starlet (wavelet) filter and a concatenation of Gaussian filters. In both cases  peak counts perform better than the power spectrum on the set of parameters [$M_{\nu}$, $\Omega_{\rm m}$, $A_{\rm s}$] respectively by 63$\%$, 40$\%$ and 72$\%$ when using a starlet filter and by 70$\%$, 40$\%$ and 77$\%$ when using a multi-scale Gaussian. More importantly, we show that when using a multi-scale approach, joining power spectrum and peaks does not add any relevant information over considering just the peaks alone. While both multi-scale filters behave similarly, we find that with the starlet filter the majority of the information in the data covariance matrix is encoded in the diagonal elements; this can be an advantage when inverting the matrix, speeding up the numerical implementation. For the starlet case, we further identify the minimum resolution required to obtain constraints comparable to those achievable with the full wavelet decomposition and we show that the information contained in the coarse-scale map cannot be neglected.

\end{abstract}



\maketitle

\section{Introduction}\label{sec:intro}

The presence of massive neutrinos affects the background evolution of the Universe as well as the evolution of cosmological perturbations and structure formation \citep{2006PhR...429..307L}. \noindent Constraining the value of the sum of neutrino masses is one of the key science goals of modern cosmology. This is not only an interesting goal per se, but it is also worth exploring because in the presence of massive neutrinos, modified gravity models may mimic the standard cosmological ($\Lambda$CDM) model, as discussed in \citep{PhysRevD.100.023508,refId0,10.1093/mnras/stz972}. This is due to the fact that massive neutrinos modify structure formation, typically reducing clustering, and can therefore allow for larger non-standard couplings than in the absence of massive neutrinos. Being able to measure massive neutrinos can also allow us to disentangle $\Lambda$CDM from alternative scenarios. From neutrino oscillation experiments \citep{CAPOZZI2016218}, we only have information about the difference of the masses squared. Hence, to fix a scale for the neutrino masses, it is necessary to assume a mass hierarchy: for a \textit{normal} hierarchy $(i.e. \,\, m_1 < m_2 < m_3)$ the lower bound on the sum of neutrino masses $M_{\nu} \equiv \sum_{\nu} m_{\nu}$ is currently predicted to be $M_{\nu} > 0.06$ $eV$ while for an \textit{inverted} hierarchy $( i.e. \,\, m_3 < m_1 < m_2)$ $M_{\nu} > 0.1$ $eV$ \citep{2019arXiv190410206D}. The latest results on the upper bound have been obtained by combining the Cosmic Microwave Background (CMB) temperature fluctuation data with CMB lensing and Baryon Acoustic Oscillations (BAO), leading to a constraint of $M_{\nu} < 0.12$ $eV$ at $95$\% confidence level \citep{Planck2018}.

Weak gravitational lensing by large-scale structure has proven to be a powerful tool to achieve constraints on cosmological parameters and its importance to precision cosmology is borne out in the scientific results of galaxy surveys such as the Canada-France-Hawaii Telescope Lensing Survey (CFHTLenS) \citep{10.1111/j.1365-2966.2012.21952.x}, the Kilo-Degree Survey (KiDS) \citep{heymans2020kids1000}, the Dark Energy Survey (DES) \citep{PhysRevD.98.043526} and Hyper SuprimeCam (HSC) \citep{10.1093/pasj/psx130, 10.1093/pasj/psz010}. In particular, it encodes the evolution of structure growth under the influence of massive neutrinos, representing a powerful tool to explore these effects and extract the corresponding cosmological information. Future galaxy surveys like \Euclid \citep{laureijs2011euclid} will be sensitive to the properties of weakly interacting particles in the eV mass range, such as massive neutrinos, and will use weak lensing as a cosmological probe to test different models and improve our knowledge of cosmological parameters. 

Moreover, in recent years, it has been shown that weak-lensing statistics higher than second order can help break degeneracies, as they take into account the non-Gaussian information encoded by the non-linear process of structure formation, such as the bispectrum \citep{10.1093/mnras/stz2862,kayo2013cosmological}, Minkowski functionals \citep{PhysRevD.88.123002,PhysRevD.85.103513}, and peak counts \citep{Kacprzak2016,Linc2015kilb,Peel2017Linc,Martinet2015,Shan2017,MartinetSchneider,Fluri_2018,zurcher2020cosmological}. In this context, we perform Bayesian inference to derive cosmological constraints on the sum of neutrino masses $M_{\nu}$, the matter density parameter $\Omega_{\rm m}$, and the primordial power spectrum amplitude $A_{\rm s}$, for a survey with \Euclid -like noise. We use as synthetic data the lensing convergence maps from \url{MassiveNus} simulations \citep{PETRI201673, Liu_2018}. Using the same suite of simulations, \citep{PhysRevD.99.083508}, \citep{Marques_2019}, and \citep{Coulton_2019} have already shown for a LSST-like survey \citep{lsstsciencecollaboration2009lsst} that combining the lensing power spectrum with higher-order statistics can provide tighter constraints on parameters. For this purpose, we perform our analysis using the lensing power spectrum and peak counts as summary statistics following \citep{PhysRevD.99.063527}. We extend the study by considering a survey with \Euclid-like noise, and to smooth the noisy convergence maps we employ a multi-scale approach, investigating a concatenation of Gaussian filters and separately a
starlet filter \citep{starck_murtagh_fadili_2010}, which was shown to be a powerful tool in the context of weak-lensing peak counts by \citep{Linc2016}.

The paper is organised as follows: Sec. \ref{sec:methodology} describes the theoretical framework of weak gravitational lensing useful for the paper and the simulations we use. Then, we illustrate the survey and noise settings, the filtering techniques that we employ for the comparison and the details of the summary statistics. In Sec. \ref{sec:analysis} we describe the interpolation method implemented to build the likelihood, the covariance matrices, the results estimators that we employ to quantify our results and the settings of the MCMC. The cosmological parameter constraints are shown in Sec. \ref{subsec:results}. We conclude in Sec. \ref{subsec:Conclusions}. 

\section{Methodology}\label{sec:methodology}
\subsection{Weak lensing}\label{subsec:weak_lensing}
The effect of gravitational lensing at comoving angular distance $f_K(\chi)$ can be described by the \textit{lensing potential}

\begin{equation}\label{Lensing_Potential}
\psi(\vec{\theta},\chi)\equiv \frac{2}{c^2}\int_0^{\chi}\mathrm{d}\chi'\frac{f_{K}(\chi - \chi')}{f_K(\chi)f_K(\chi')}\Phi(f_K(\chi')\vec{\theta},\chi') \,\, ,
\end{equation}

\noindent which defines how much the gravitational potential $\Phi$  arising from a mass distribution changes the direction of a light path. In this expression $K$ is the spatial curvature constant of the universe, $\chi$ is the comoving radial coordinate, $\vec{\theta}$ is the angle of observation, and $c$ is the speed of light. 
As we are in $\Lambda$CDM, the two Bardeen gravitational potentials are here assumed to be equal and the metric signature is defined as $(+1,-1,-1,-1)$.
In particular, under the \textit{Born approximation} the effect of the lensing potential on the shapes of background galaxies in the weak regime can be summarised by its variation with respect to $\vec{\theta}$. Formally, this effect can be described by the elements of the lensing potential Jacobi matrix:

\begin{equation}\label{Lensing_Jacobi_Matrix_elements}
A_{ij}=\delta_{ij}-\partial_i \partial_j \psi,
\end{equation}

which can be parametrised as

\begin{equation}\label{Lensing_Jacobi_Matrix}
A=\left (
\begin{array}{cc}
1-\kappa -\gamma_1 &  -\gamma_2   \\
 -\gamma_2  & 1-\kappa +\gamma_1 \\
\end{array}
\right ),
\end{equation}

where $(\gamma_1, \gamma_2)$ are the components of a spin-2 field $\gamma$ called \textit{shear}, and $\kappa$ is a scalar quantity called \textit{convergence}. They describe respectively the anisotropic stretching and the isotropic magnification of the galaxy shape when light passes through large-scale structure. Equation \eqref{Lensing_Jacobi_Matrix_elements} and Eq. \eqref{Lensing_Jacobi_Matrix} define the shear and the convergence fields as second-order derivatives of the lensing potential:
\begin{equation}\label{shear}
\gamma_1\equiv\frac{1}{2}(\partial_1 \partial_1 - \partial_2 \partial_2)\psi\quad \gamma_2\equiv \partial_1\partial_2\psi
\end{equation}
\begin{equation}\label{convergence}
 \kappa\equiv\frac{1}{2}(\partial_1 \partial_1 + \partial_2 \partial_2)\psi=\frac{1}{2}\nabla^2\psi.
\end{equation}
The weak-lensing field is a powerful tool for cosmological inference. The shear is more closely related to actual observables (i.e., galaxy shapes), while the convergence, as a scalar field, can be more directly understood in terms of the matter density distribution along the line of sight. This can be seen by inserting the lensing potential defined in Eq. \eqref{Lensing_Potential} inside Eq. \eqref{convergence} and using the fact that the gravitational potential $\Phi$ is related to the matter density contrast $\delta=\Delta\rho/ \bar{\rho}$ through the Poisson equation $\nabla^2 \Phi=4\pi Ga^2 \bar{\rho}\delta$. Expressing the mean matter density in terms of the critical density $\rho_{c,0}=3H_0^2/(8\pi G)$, the convergence field can be rewritten as
\begin{equation}\label{Convergence}
\kappa(\vec{\theta})=\frac{3H_0^2\Omega_{\rm m}}{2c^2}\int_0^{\chi_\mathrm{lim}}\frac{\mathrm{d}\chi}{a(\chi)}g(\chi)f_{K}(\chi)\delta(f_K(\chi)\vec{\theta},\chi),
\end{equation}
 
\noindent where $H_0$ is the Hubble parameter at its present value, and
\begin{equation}
g(\chi) \equiv \int_{\chi}^{\chi_\mathrm{lim}} \mathrm{d}\chi' n(\chi') \frac{f_K(\chi'-\chi)}{f_K(\chi')}
\end{equation}

\noindent is the \textit{lensing efficiency}.
\noindent Equation \eqref{Convergence} relates the convergence $\kappa$ to the 3D matter overdensity field $\delta(f_K(\chi)\vec{\theta},\chi)$, and it describes how the lensing effect on the matter density distribution is quantified by the lensing strength at a distance $\chi$ that directly depends on the normalised source galaxy distribution $n(z)\mathrm{d}z=n(\chi)\mathrm{d}\chi$ and on the geometry of the universe through $f_K(\chi)$ along the line of sight. For a complete derivation see \citep{Kilbinger_2015} and \citep{1992grle.book.....S}. 

\subsection{Simulations}\label{subsec:simulations}
In this paper, we use the Cosmological Massive Neutrino Simulations (\url{MassiveNus}), a suite of publicly available N-body simulations released by the Columbia Lensing group \citep{columbia_lensing}. It contains 101 different cosmological models obtained by varying the sum of neutrino masses $M_{\nu}$, the total matter density parameter $\Omega_{\rm m}$ and the primordial power spectrum amplitude $A_{\rm s}$ at the pivot scale $k_0=0.05$ Mpc$^{-1}$,  in the range 
$M_{\nu}=[0, 0.62]$ eV, $\Omega_{\rm m} =[$0.18, 0.42$]$ and $A_{\rm s}\cdot 10^{9}=[$1.29,  2.91$]$. The reduced Hubble constant $h=0.7$, the spectral index $n_s=0.97$, the baryon density parameter $\Omega_b = 0.046$ and the dark energy equation of state parameter $w=-1$ are kept fixed under the assumption of a flat universe. The fiducial model is set at $ \left\lbrace M_{\nu}, \Omega_{\rm m}, 10^{9}A_{\rm s}  \right\rbrace $=$ \left\lbrace 0.1, 0.3, 2.1 \right\rbrace $.

The presence of massive neutrinos is taken into account assuming normal hierarchy and using a linear response method, where the evolution of neutrinos is described by linear perturbation theory but the clustering occurs in a non-linear cold dark matter potential. The simulations have a 512 Mpc/h box size with $1024^3$ CDM particles. They are implemented using a modified version of the public tree-Particle Mesh (tree-PM) code \url{Gadget2} with a neutrino patch, describing the impact of massive neutrinos on the growth of structures up to $k = 10$ h Mpc$^{-1}$. For a complete description of the implementation and the products see  \citep{Liu_2018}. We use the simulated convergence maps as mock data for our analysis. When dealing with real data the actual observable is the shear field that can be converted into the convergence field following \citep{1993ApJ...404..441K}. We bypass this step from $\gamma$ to $\kappa$ and work with the convergence maps directly provided as products from \url{MassiveNus}.
The maps are generated using the public ray-tracing package \url{LensTools} \citep{PETRI201673} for each of the 101 cosmological models at five source redshifts $z_s=\left\lbrace 0.5, 1.0, 1.5, 2.0, 2.5 \right\rbrace $. Each redshift has 10000 different map realisations obtained by rotating and shifting the spatial planes. Each $\kappa$ map has $512^2$ pixels, corresponding to a $12.25$ deg$^{2}$ total angular size area in the range $\ell \in [100 , 37000]$ with a resolution of 0.4 arcmin. 

\subsection{Noise and survey specifications}\label{Subsec:Noise_def}
The method described in this paper can be applied to any given survey. For illustration purposes, we perform here a tomographic study using redshifts $z_s=\left\lbrace 0.5, 1.0, 1.5, 2.0 \right\rbrace$ and mimicking the noise expected for a survey like \Euclid \citep{laureijs2011euclid, EuclidForecast2020}. Specifically, at each source redshift we produce 10000 map realisations of Gaussian noise with mean zero and variance
\begin{equation}\label{eq:noise}
\sigma_{n}^{2}=\frac{\langle \sigma_{\epsilon}^2\rangle}{n_\mathrm{gal}A_\mathrm{pix}} \,\, ,
\end{equation}
\noindent where we set the dispersion of the ellipticity distribution to $\sigma_{\epsilon}=0.3$, and the pixel area is given by $A_\mathrm{pix} \simeq 0.16$ arcmin$^2$. The redshift dependence that makes a tomographic investigation possible is encoded in the source galaxy redshift distribution, for which we assume the parametric form
\begin{equation} \label{n_z_pure}
n(z)= \mathcal{C} \left( \frac{z}{z_0} \right)^{\alpha} \exp{\left[- \left( \frac{z}{z_0} \right)^{\beta}\right]},
\end{equation}
with $\alpha=2$, $\beta=3/2$ $z_0=0.9/ \sqrt{2}$ as in \citep{laureijs2011euclid, EuclidForecast2020}, and $\mathcal{C}$ is the normalization constant to guarantee the constraint $\int _{z_{\min}}^{z_{\max}} n(z)$ $ \mathrm{d}z = 30 $ arcmin$^{-2}$. Then, we compute the galaxy number density at each bin as
\begin{equation}\label{eq:n_bin}
n^{i}_\mathrm{gal}=\mathcal{C}  \int_{z_i^{-}}^{z_i^{+}} n(z) \mathrm{d}z,
\end{equation}
where $z_i^{-}, z_i^{+}$ are the edges of the $i^{th}$ bin. We adapt the binning choice to the provided simulation settings, assuming that we observe galaxies within a small range around the actual source redshift. This leads to the values for the galaxy number density $n_\mathrm{gal}$ per source redshift bin $z_s$ provided in \autoref{tab:redshifts_n_gal}:  

\begin{table}[h!]
\begin{tabular}{ccccc}
\hline
\hline
$\mathbf{z_s}$&0.5&1.0&1.5&2.0 \\ \cline{1-5}
$\mathbf{n_\mathrm{gal}}$&  11.02&  11.90&  5.45& 1.45 \\ 
\cline{1-5}
\end{tabular}
\caption{\label{tab:redshifts_n_gal}Values of $n_{\text{gal}}$ for each source redshift $z_s$. We adapt the binning choice to the provided simulation settings, assuming that we observe galaxies within a small range around the actual redshift. In practice, this means considering as  bin edges $\left\lbrace {0.001, 0.75, 1.25, 1.75, 2.25} \right\rbrace$, in order to compute the integral in Eq. \eqref{eq:n_bin}.}
\end{table}

\subsection{Gaussian and starlet filters}\label{subsec:filters}

\begin{figure*}
\includegraphics[width=\textwidth]{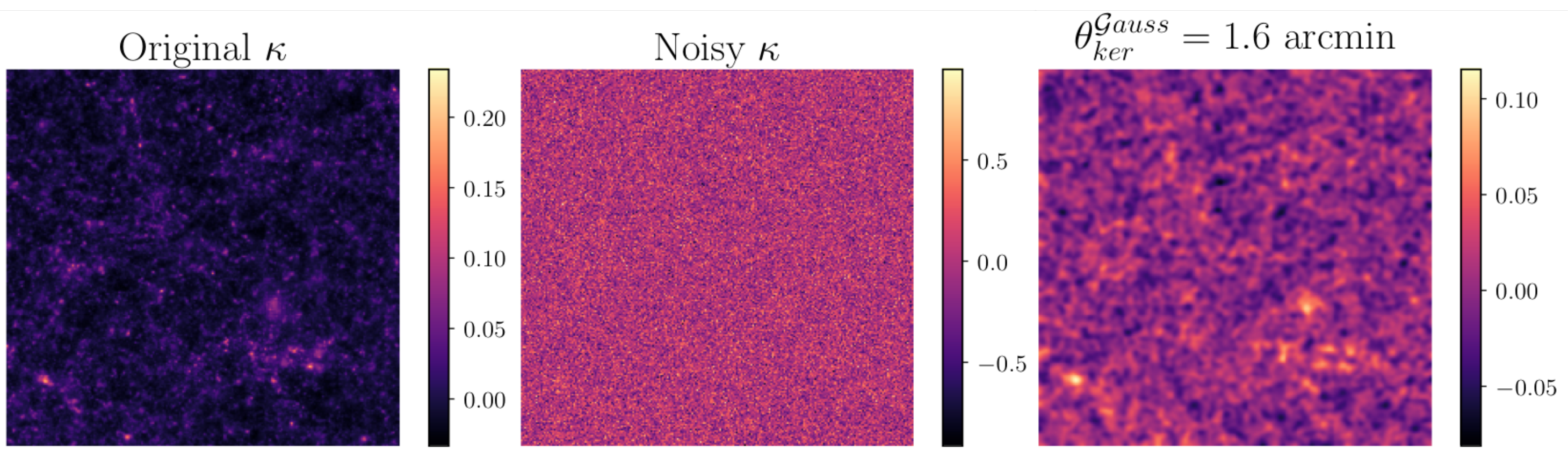}
\caption{\label{Filters}Convergence maps $\kappa$ are noiseless. We apply Gaussian noise and then filter the map using either the Gaussian or starlet filtering. For illustration purposes, we show the Gaussian filtering with $\theta_\mathrm{ker}^{\mathcal{G}auss}=1.6$ arcmin of one map realisation for the fiducial model $ \left\lbrace M_{\nu}, \Omega_{\rm m}, 10^{9}A_{\rm s}  \right\rbrace $=$ \left\lbrace 0.1, 0.3, 2.1 \right\rbrace $. The color bar on the right of each map describes values of the convergence field $\kappa$. For each realisation of the 10000 maps provided for each redshift we generate 10000 noise maps as described in Sec. \ref{Subsec:Noise_def} corresponding to the different value of $n_\mathrm{gal}$ respectively for $z_s=[0.5, 1.0, 1.5, 2.0]$.}
\end{figure*}

In order to access the signal in the convergence maps at small scales, where they are mostly dominated by noise, we filter them, considering a multi-scale analysis compared to a single-scale analysis. First, we use a single Gaussian kernel of size $\theta_\mathrm{ker}$, defined as 
\begin{equation}\label{eq:Gaussian_kernel}
\mathcal{G}(\theta;\theta_\mathrm{ker})=\frac{1}{\sqrt{2 \pi} \theta_\mathrm{ker}} e^{-\theta^2/(2\theta_\mathrm{ker}^2)} \,\, ,
\end{equation}
which was also used in e.g. \cite{PhysRevD.99.063527, 10.1093/mnras/staa1098}. We then compare results with those obtained when applying instead a concatenation of Gaussian filters and an \textit{Isotropic Undecimated Wavelet Transform}, also known as a starlet transform \citep{4060954}, which allows us to represent an image $I$ as a sum of wavelet coefficient images $w_j$ and a coarse resolution image $c_{J}$.
\begin{figure}[h!]
\includegraphics[width=\columnwidth]{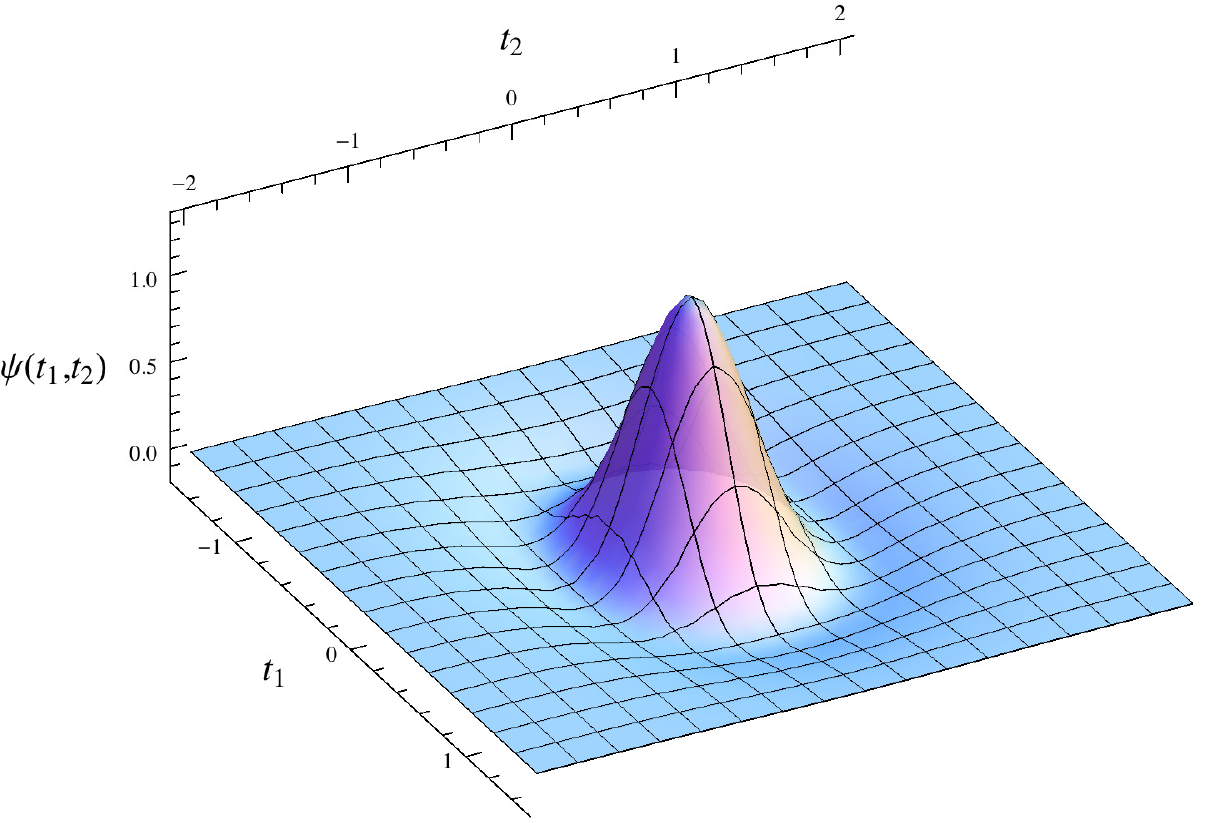}
\includegraphics[width=\columnwidth]{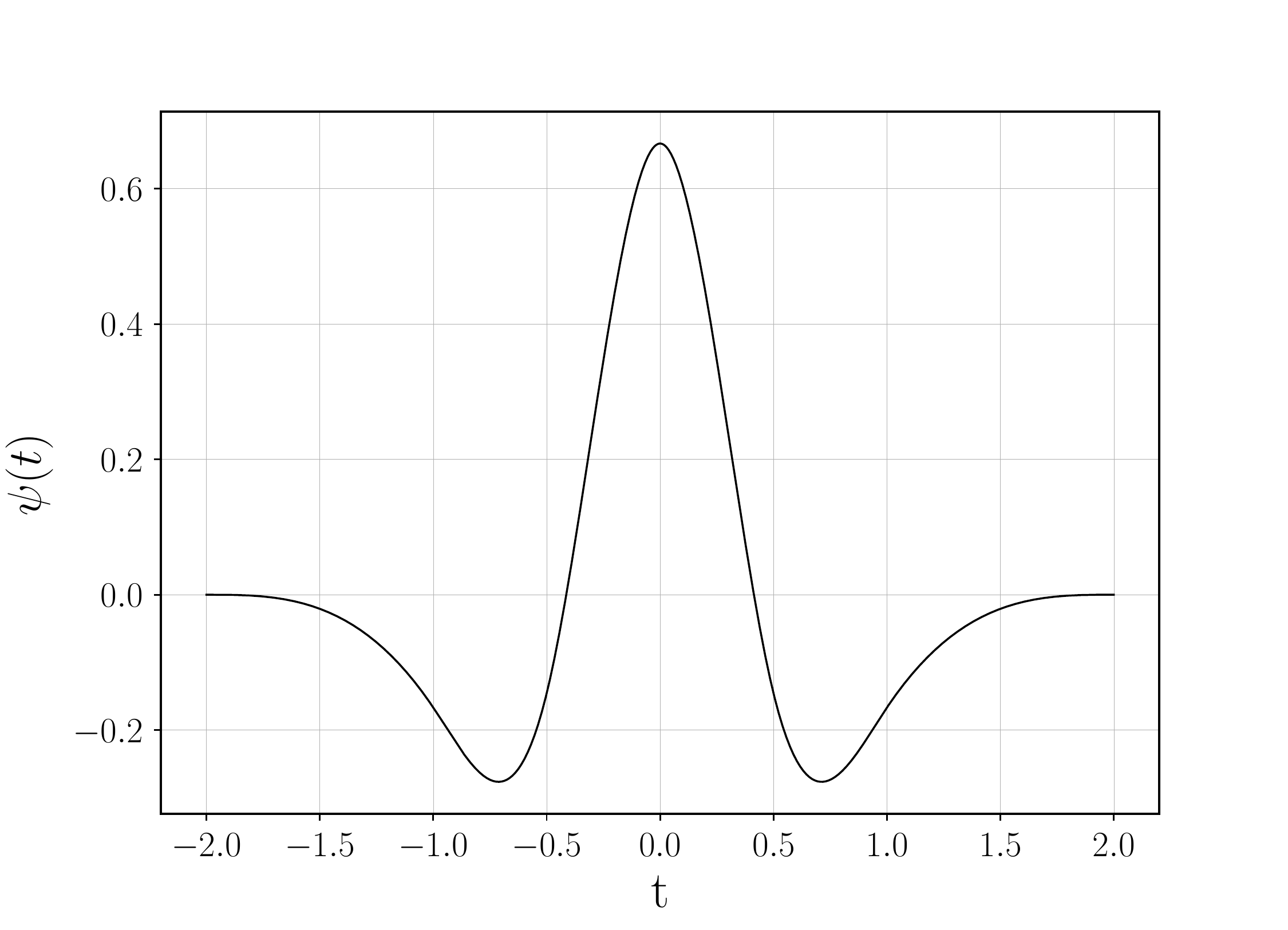}
\caption{\label{Starlet_1D_2D_profiles}We show the 2D starlet function (top panel) as defined in Eq. \eqref{Starlet_equation} and its 1D profile (bottom panel). Being a wavelet it is a compensated function, i.e. it integrates to zero over its domain. This comes from the \textit{admissibility condition} for the wavelet function $\psi$: $\int_{0}^{+ \infty} | \hat{\psi(k)} |^{2} \frac{\mathrm{d}k}{k} < + \infty$ which implies that $\int \psi(x) \mathrm{d}x =0 $ and it has compact support in $[-2,2]\times[-2,2]$. Its shape emphasises round features, making it very efficient when dealing with peaks.\label{starlet_function}}
\end{figure} 
The starlet filter is a wavelet transform, i.e. a function satisfying the \textit{admissibility condition} that allows for the simultaneous processing of data at different scales. An original map \textit{I} is decomposed by this transform into a coarse version of it $c_{J}$ plus several images of the same size at different resolution scales \textit{j}:
\begin{equation}\label{eq:Starlet_decomposition}
I(x,y)=c_{J}(x,y) + \sum_{j=1}^{j_{\max}}w_j(x,y),
\end{equation}
where wavelet images $w_j$ represent the details of the original image at dyadic (powers of two) scales corresponding to a spatial size of $2^j$ pixels and $J=j_{\max}+1$. The starlet wavelet function $\psi$ is derived from a B-spline function $\phi$ of order 3:
\begin{equation}\label{Starlet_equation}
\psi(t_1,t_2)=4 \phi(2t_1,2t_2)-\phi(t_1,t_2)
\end{equation}
with 
\begin{equation}
\phi (t) = \frac{1}{12}(|t-2|^3-4|t-1|^3+6|t|^3-4|t+1|^3+|t+2|^3)   
\end{equation}
and $\phi(t,t')=\phi(t)\phi(t')$. For a complete description and derivation of the starlet transform algorithm, see  \citep{starck_murtagh_fadili_2010}. We show its 1D and 2D profiles in Fig. \ref{Starlet_1D_2D_profiles}. One of the advantages of employing a starlet filter is provided by its multi-scale analysis, namely its ability to investigate and extract the information encoded at different scales at the same time \citep{starck_murtagh_bijaoui_1998}.
Hence, the starlet transform presents the properties to compute efficiently $J$ scales with a fast algorithm with a complexity of $\mathcal{O}(N^2 \log{N})$ for an image of $N \times N$ pixels and to analyse data with compensated aperture filters with finite support. See also \citep{Peel2017Linc, Peel2018} for further details on the advantages of wavelet starlet analysis. \noindent The following example illustrates how we can compare results from these two different filtering schemes. Applying a starlet transform with $j_{max}=4$ to a map with 0.4 arcmin pixel size will result in a decomposition of four maps with resolutions [0.8, 1.6, 3.2, 6.4] arcmin plus the coarse-scale map. For our study, we will consider as finest scale $\theta^{\mathcal{S}t}_{ker}=1.6$ arcmin, being a more realistic choice in terms of resolution for convergence maps coming from \Euclid -like survey data. We will therefore focus on the set of scales [1.6, 3.2, 6.4] arcmin plus the coarse map.
Concerning the multi-Gaussian filters, to fairly compare them to starlets, we set the standard deviations of the Gaussian filters such that their maximum matches that of the corresponding single starlet scale profile, resulting in a concatenation of Gaussians respectively with $\theta^{\mathcal{G}}_{ker}=[1.2, 2.7, 5.5, 9.5]$ arcmin. Based on the above, in our study we compare cosmological constraints obtained using noisy maps smoothed from a single-Gaussian kernel with the ones obtained from a multi-Gaussian analysis and from a starlet decomposition. We exclude the observables corresponding to $0.8$ arcmin in our analysis after having verified that this does not cost any loss of information. The starlet transform can be seen as multi-Gaussian filtering where each Gaussian kernel is replaced by a compensated filter. In Fig. \ref{Filters} we show the result of the filtering procedure for a Gaussian kernel: given the original convergence map $\kappa$, we add white noise as described in Sec. \ref{Subsec:Noise_def} and then we filter the noisy map with the chosen kernel. To  extract and investigate the cosmological information encoded in the weak lensing convergence maps, we compute the power spectrum (PS) and peak counts as summary statistics.

\subsubsection{Convergence power spectrum}
\noindent To provide a statistical estimate of the distribution of the convergence field, the first non-zero order is given by its second moment, which is commonly described by the \textit{two-point correlation function} ($2$PCF) in real space $\langle \kappa(\theta)\kappa(\theta') \rangle$, or by its counterpart in Fourier space, the \textit{convergence power spectrum}:

\begin{equation}
C_{\kappa}(\ell)=\frac{9\Omega_{\rm m}^2 H_0^4 }{4c^4} \int_0^{\chi_\mathrm{lim}} \mathrm{d}\chi \frac{g^2(\chi)}{a^2(\chi)} P_{\delta} \left(\frac{\ell}{f_{\kappa}(\chi)},\chi \right)
\end{equation}

\noindent where $P_{\delta}$ represents the 3D matter power spectrum, directly related to the matter density distribution $\delta$ in Eq. \eqref{Convergence} of the weak-lensing convergence field. In this study, we compute the power spectrum of the noisy filtered convergence maps: for a given cosmology we add Gaussian noise to each realisation of $\kappa$. For each redshift we generate a different set of noise maps following Eq. \eqref{eq:noise}.

To filter the maps we employ a Gaussian kernel with smoothing size $\theta_\mathrm{ker}=1$ arcmin and consider angular scales with logarithmically spaced bins in the range $\ell=[300,5000]$. We compute the power spectra using \url{LensTools} for each of the 10000 realisations per cosmology and then we take the average over the realisations. We parallelise our code using \url{joblib} \citep{joblib} to accelerate processing due to the large number of realisations per cosmology.

\subsubsection{Peak counts}
\begin{figure}

\includegraphics[width=\columnwidth]{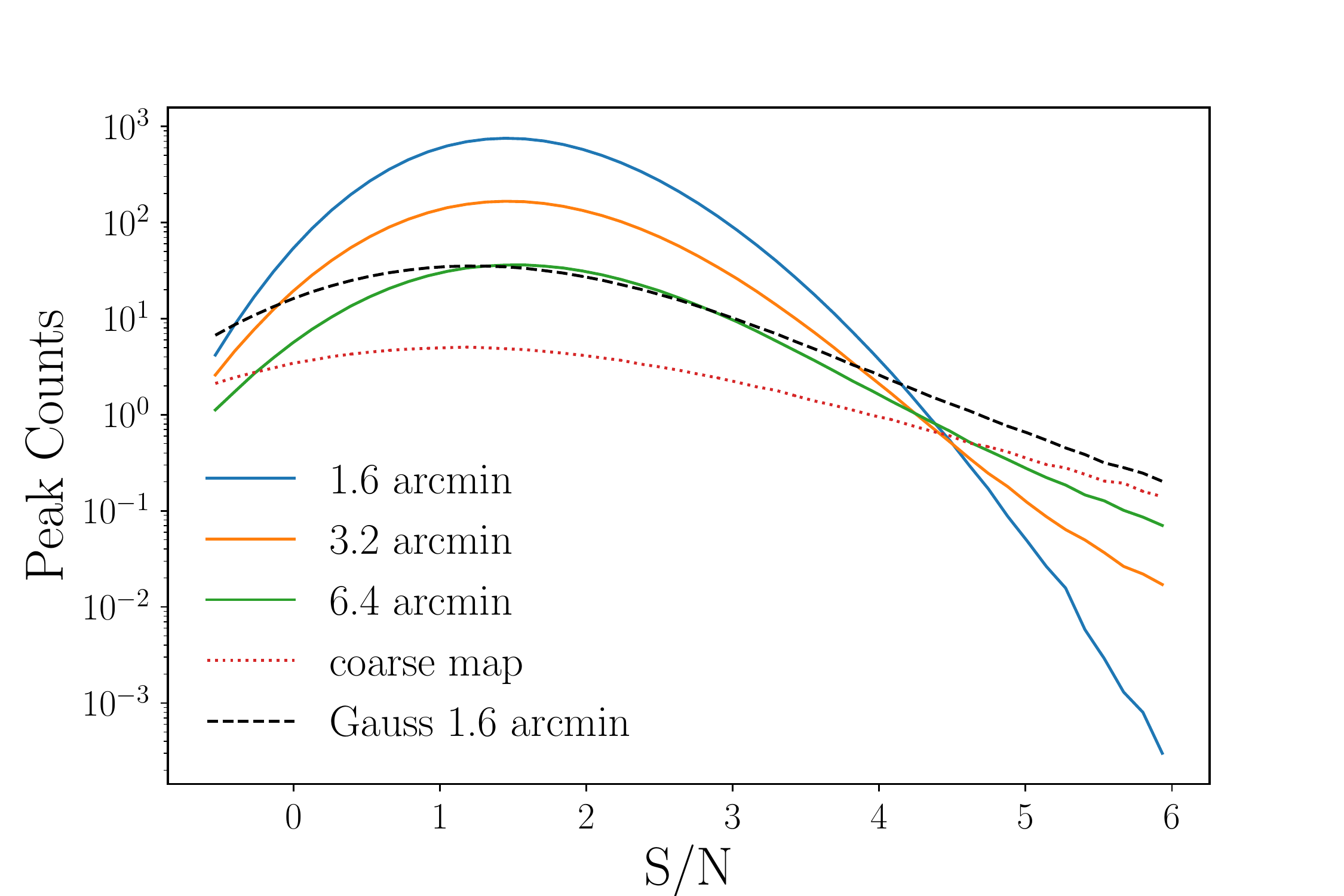}
\caption{Peak Counts distribution in logarithmic scale for each starlet scales resolutions: [1.6, 3.2, 6.4] arcmin and the coarse maps (red dotted line) and the Gaussian case (black dashed line). Due to the decomposition at different scales, for each map filtered with the starlet there are 4 different distributions. Indeed, the number of counts depends on the resolution: the larger the smoothing size (the lower the frequency) the smaller the number of peaks.}\label{fig:starlet_scales}
\end{figure}

Second order statistics such as the power spectrum have been widely used in studies performing cosmological parameter estimation with cosmic shear; see, for example, \citep{Jee_2013, PhysRevD.94.022001, 10.1093/mnras/stw2805}. However, it is well known that it is necessary to go beyond second-order statistics in order not to lose the non-Gaussian information in the matter distribution due to the weak-lensing correlations arising in the non-linear regime. Recently, several studies have considered weak-lensing peak counts as a robust and complementary probe to the power spectrum to constrain cosmological parameters \citep{Kacprzak2016, Linc2015kilb,Peel2017Linc,Martinet2015,Shan2017,MartinetSchneider,Fluri_2018, PhysRevD.99.063527, Linc2016}. The physical meaning of weak lensing peaks can be identified in the fact that they trace regions where the value of the convergence field is high, hence, they are in some way associated to massive structures. Nevertheless, their exact relation with halos is not trivial due to projection and noise that can generate false detections. In this paper, we detect and count weak lensing peaks on the noisy filtered maps using our own code \citep{lenspack}. We compute peaks as local maxima of the signal-to-noise field $\nu$ i.e. as a pixel of larger value than its eight neighbors in the image. We define the signal to noise field $\nu=S/N$ as the ratio between the noisy convergence $\kappa$ convolved with the filter $\mathcal{W}(\theta_\mathrm{ker})$ over the smoothed standard deviation of the noise for each realisation per redshift: 
\begin{equation}\label{eq:singal_to_noise}
\nu=\frac{(\mathcal{W} \ast \kappa )(\theta_\mathrm{ker})}{\sigma_{n}^\mathrm{filt}},
\end{equation}

where $\mathcal{W}(\theta_\mathrm{ker})$ can be the single-Gaussian, the multi-Gaussian or the starlet filter. Concerning $\sigma_{n}^\mathrm{filt}$, its definition depends on the employed filter. For a Gaussian kernel it is given by the standard deviation of the smoothed noise maps,
while for the starlet case we need to estimate the noise at each wavelet scale for each image per redshift. To estimate the noise level at each starlet scale we follow  \citep{Starck_1998} and use the fact that the standard deviation of the noise at the scale $j$ is given by $\sigma_j=\sigma^e_j\sigma_I$, where $\sigma_I$ is the standard deviation of the noise of the image and $\sigma^e_j$ are the coefficients obtained by taking the standard deviation of the starlet transform of a Gaussian distribution with standard deviation one at each scale $j$. To estimate $\sigma_I$ we take the \textit{median absolute deviation}\footnote{For a Gaussian distribution the \textit{median absolute deviation} (MAD) and the standard deviation are directly related as: MAD$/\sigma=0.6745$. We choose to use this estimator since it is more robust when dealing with non-normal distributions (being more resistant to outliers in a data set) to have a more general implementation in our pipeline.} of the noisy convergence map. We do this for each one of the 10000 realisations for each cosmology and then take the average over the realisations.
 We consider the peak distribution for 41 linearly spaced bins within the range $\nu=[-0.6,6]$, based on the outcomes of the companion paper \citep{PhysRevD.99.063527} where it is shown that including the low ($S/N<1$), medium ($1<S/N<3$) and high peaks ($S/N>3$) jointly give the best constraints. Moreover, low and medium peaks, typically formed due to multiple much smaller halos than the single halos that cause the high peaks \citep{PhysRevD.94.043533}, contain a similar level of information as the high peaks. In Fig. \ref{fig:starlet_scales} we show for illustration purposes the peak counts distribution in logarithmic scale for each starlet scale and for the Gaussian filter case. We see that the number of counts depends on the resolution: the larger the smoothing size (the lower the frequency) the smaller the number of peaks. We have investigated the impact of the binning on the results by testing different boundaries, and we have found that choosing 41 bins instead of 50 decreases the condition number of the data matrix as shown in Sec. \ref{subsec:Matrix_condition_number}, hence facilitating its inversion during the likelihood analysis. We have also considered the minimum and the maximum values of the $S/N$ maps as bin edges, and we have seen that this choice is not very convenient, since it
 increases the condition number by two orders of magnitude.

\section{Analysis}\label{sec:analysis}
\subsection{Likelihood}\label{subsec:Gaussian_Processes}
To perform Bayesian inference and get the probability distributions of the cosmological parameters, we use a Gaussian likelihood for a cosmology-independent covariance:
\begin{equation}\label{Likelihood_function}
\log \mathcal{L}(\mathbf{\theta})=\frac{1}{2}(d-\mu(\mathbf{\theta}))^{T}C^{-1}(d-\mu(\mathbf{\theta})),
\end{equation}

\noindent where $d$ is the data array, $C$ is the covariance matrix of the observable, $\mu$ the expected theoretical prediction as a function of the cosmological parameters $\theta$. In our case, the data array is the mean over the (simulated) realisations of the power spectrum or peak counts or combination of the two for our fiducial model. Cosmological parameters are the ones for which simulations are available, namely $\lbrace{  M_{\nu},\Omega_{\rm m},A_{\rm s}}\rbrace$.

In order to determine the relation between the observable and the models $\mu(\mathbf{\theta})$, i.e. to be able to have a prediction of the power spectrum and the peak counts given a new set of cosmological parameters $\lbrace{  M_{\nu},\Omega_{\rm m},A_{\rm s}}\rbrace$, we employ an interpolation with Gaussian Processes Regression (GPR, \citep{Rasmussen:2005:GPM:1162254}) using the \url{scikit-learn} python package. Gaussian Processes are a generic supervised learning method that, via an assumption of smoothness between parameters with close values, allows one to compute the prediction for an observable at a new given point in parameter space. The cosmological parameters and the corresponding observables (power spectrum and peak counts or the two statistics combined) from the simulations are used as a training set, i.e. as the input for the GPR. Then, the Gaussian Processes act by assuming that for a new point in parameter space $\theta_*$ which is sufficiently close to a known point $\theta$ belonging to the training set, the corresponding observable will be described by a joint normal distribution along with the known observable. This can be summarised by:
\begin{equation*}
\begin{bmatrix} 
f \\ 
f_*
\end{bmatrix}\sim
\mathcal{N}
\begin{pmatrix} 
\begin{bmatrix} 
\mu \\ 
\mu_*
\end{bmatrix} , \begin{bmatrix} 
K(\theta,\theta)+\sigma_n^2I & K(\theta,\theta_*)  \\ 
K(\theta_*,\theta) & K(\theta_*,\theta_*)
\end{bmatrix}
\end{pmatrix},
\end{equation*}

\noindent where $K(\theta,\theta')$ is the \textit{kernel} of the Gaussian processes that assesses the smooth relation among points in parameter space and has form of an \textit{anisotropic squared exponential} function. $\sigma_n$ is the standard error of the noise level in the targets, namely in our case the noise given by the fact that we take the mean over 10000 realisations for each observable and each bin. More specifically, for each bin we compute the observable mean and its corresponding standard error over the 10000 realisations available from the simulations. The fitting procedure takes then as input the three cosmological parameters and the re-scaled mean of the observable for each bin, while the standard errors are added as dual coefficients of the training data points in kernel space, i.e. as a regularization term to the diagonal of the kernel matrix to take into account the noise level on the mean. This results in a number of GP corresponding to the number of bins that are then taken in by a prediction function. The latter reads new points in parameter space and returns the corresponding observable predictive distribution (power spectrum, peaks or the two statistics jointly) and its standard deviation. The hyperparameters are fit with the standard marginal likelihood approach. For the validation we compare the prediction of a given model obtained with the GP emulator excluding the corresponding model from the simulation, for 10 models near the fiducial model following \citep{PhysRevD.99.083508}. We find differences at the sub-percent level that always lie within the statistical error consistent with \citep{Marques_2019, Coulton_2019, PhysRevD.99.063527}.

\subsection{Covariance matrices}\label{Section:covariance}
We use the independent fiducial massless neutrino simulation, defined by $ \left\lbrace M_{\nu}, \Omega_{\rm m}, 10^{9}A_{\rm s}  \right\rbrace $=$ \left\lbrace 0.0, 0.3, 2.1 \right\rbrace $ and obtained from initial conditions different from the massive simulations to compute the covariance matrices of the data. We consider a parameter-independent covariance to reduce the risk of assigning an excess of information to the observables in the context of a Gaussian likelihood assumption, following the results of \citep{2013A&A...551A..88C}. The covariance matrix elements are computed as

\begin{equation}\label{covariance_element}
C_{ij}=\sum\limits_{r=1}^{N} \frac{(x_{i}^{r} - \mu_i)(x_{j}^{r} - \mu_j)}{N-1}
\end{equation}

\noindent where $N$ is the number of observations (in this case the 10000 realizations), $x_i^{r}$ is the value of the power spectrum or the peak counts in the $i^{th}$ bin for a given realisation $r$ and 
\begin{equation}
\mu_i=\frac{1}{N}\sum_r x_{i}^{r}
\end{equation}

\noindent is the mean of the power spectrum or the peak counts in a given bin over all the realisations. 

\begin{figure*}
\includegraphics[width=\textwidth]{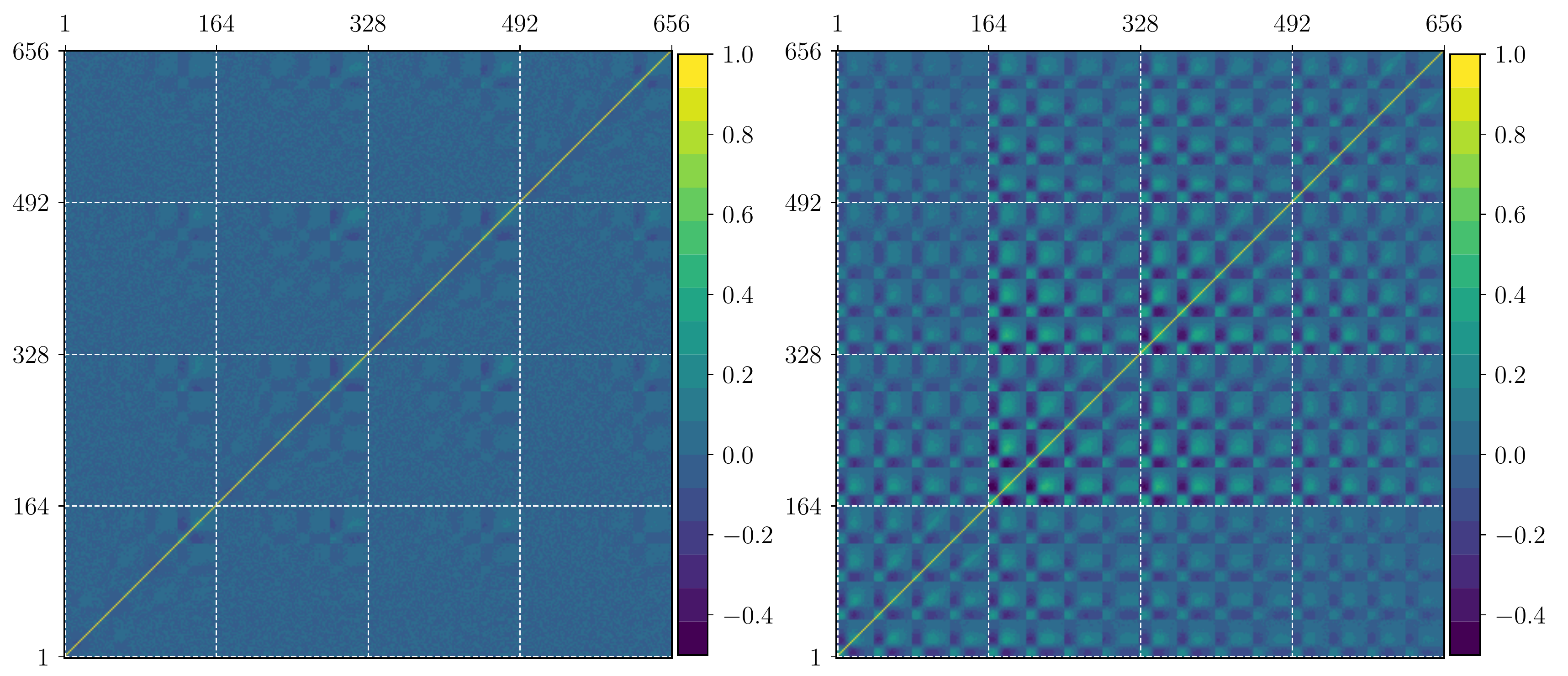}
\caption{We show the correlation coefficients respectively for the starlet (left) and the multi-Gaussian (right) peak counts. The white dashed lines split the contribution of the different redshifts: the bin range [1,164] refers to the correlations at $z_s=0.5$, the bin range [165, 328] refers to the correlations at $z_s=1.0$, the bin range [328, 492] refers to the correlations at $z_s=1.5$ and the bin range [493, 656] refers to the correlations at $z_s=2.0$. Each redshift contribution is split in the four different scales (the four mini-blocks inside the white boxes) in increasing order, i.e. [1.6, 3.2, 6.4, coarse] arcmin for the starlet and [1.2, 2.7, 5.5, 9.5] arcmin for the multi-Gaussian, where each scale is binned in 41 values of $S/N$ in the range $[-0.6, 6]$. We notice how the starlet decomposition has a tendency to diagonalise the observable correlation matrix while the off-diagonal terms of the multi-Gaussian matrix show more correlations along the scales and among $S/N$ values in the mini-blocks. \label{fig:Covariances}}
\end{figure*}

In Fig. \ref{fig:Covariances} we show the correlation coefficients of the multi-scale tomographic peak counts: in the left panel the starlet case and in the right panel the multi-Gaussian case. The matrices are organised as follows. From left to right, the four main blocks are for the tomographic redshifts respectively in the order $z_s=[0.5, 1.0, 1.5, 2.0]$. Within each of the main blocks, there are four sub-blocks representing the filter scale, i.e. the scales [1.6', 3.2', 6.4', coarse] for the starlet and [1.2', 2.7', 5.5', 9.5'] for the multi-Gaussian. Each scale is binned in 41 values of signal to noise in the range $S/N=[-0.6, 6]$. We see that the starlet decomposition has a tendency to make the matrix more diagonal, while the off-diagonal terms for a multi-Gaussian show more correlations between the scales and for small and high values of $S/N$. Consistent with this, we notice that the most correlated bins in the starlet case are the ones corresponding to the coarse scale (the last mini-block for each of the main blocks) whose profile indeed closely mimics a Gaussian, as one can see in the last panel of Fig. \ref{fig:gaussian_starlet_functions}. Furthermore, we take into account the loss of information due to the finite number of bins and realisations by adopting for the inverse of the covariance matrix the estimator introduced by \citep{2007A&A...464..399H}: 

\begin{equation}\label{Cov_est_corrected}
C^{-1}=\frac{N-n_\mathrm{bins}-2}{N-1}C_{*}^{-1},
\end{equation}

\noindent where $N$ is the number of realisations, $n_\mathrm{bins}$ the number of bins, and $C_{*}$ the covariance matrix computed for the power spectrum and peak counts, whose elements are given by Eq. \eqref{covariance_element}. We also scale the covariance for a \Euclid sky coverage by the factor $f_\mathrm{map}/f_\mathrm{survey}$, where $f_\mathrm{map}=12.25$ deg$^{2}$ is the size of the convergence maps and $f_{Euclid}=15000$ deg$^{2}$. In using Eq. \eqref{Cov_est_corrected}, we do not expect all biases to be removed from our parameter inference, as this has already been ruled out in \citep{10.1093/mnrasl/slv190} and \citep{10.1093/mnras/stw2697}. Nevertheless, we rely on the fact that the number of realisations that we are using (10000) is sufficiently large and greater than $n_\mathrm{bins}$ to consider it a reliable estimator for our purposes\footnote{Indeed, the value of the correction coefficient is close to 1 for each analysis we perform. However, considering that the results of \citep{10.1093/mnras/stw2697} quantify the loss of information also in the case of a \Euclid-like survey, it would be worth it to reproduce our study applying their restoration technique to generalize our analysis.}.

\subsection{Result estimators}
In order to quantify our results, we use estimators common in the literature, whose definitions we recall here for convenience.

\subsubsection{Figure of Merit}

To have an approximate quantification of the size of the parameter contours that we use to compare their constraining power, we consider the following Figure of Merit (FoM) as defined in \citep{EuclidForecast2020}:
\begin{equation}\label{eq:Figure_of_Merit}
\text{FoM}=\left ({\det{(\Tilde{F})}}\right)^{1/n}
\end{equation}

\noindent where $\Tilde{F}$ is the marginalised Fisher submatrix that we estimate as the inverse of the covariance matrix among the set of cosmological parameters $\left\lbrace M_{\nu}, \Omega_{\rm m}, 10^{9}A_{\rm s}  \right\rbrace $ obtained with the MCMC chains. In the exponent, $n$ is equal to the parameter space dimensionality, e.g. $n=2$ for FoM in a 2-dimensional plane between two parameters while $n=3$ if we take the Fisher matrix among the three parameters. We show the values of the FoM four our observables in \autoref{tab: Table_FoM}.

\subsubsection{Figure of correlation}
To quantify the correlations among the parameters we use the \textit{Figure of Correlation} \citep{CASAS201773,EuclidForecast2020}:

\begin{equation}\label{Eq:fig_of_correlation}
\text{FoC}=\sqrt{\det(\bf{P}^{-1})},
\end{equation}
where $\mathbf{P}$ is the correlation matrix whose elements are defined as $P_{\alpha \beta}=C_{\alpha \beta}/\sqrt{C_{\alpha \alpha}C_{\beta \beta}}$, with $C_{\alpha \beta}$ the covariance between the cosmological parameters $\alpha$ and $\beta$ as defined in the previous section. When the parameters are fully uncorrelated FoC $=$ 1, while for FoC $>$ 1 the off-diagonal terms are non-zero, indicating an increasing presence of correlations among parameters as FoC increases. The values of the FoC for our constraints are shown in  \autoref{Figure_of_correlation} and we will comment on them in Sec. \ref{subsec:results}. 

\begin{table*}[!htbp]
\begin{tabular}{lcccc}
\hline
\hline
\textbf{Condition Number} &  \textbf{Single-Gaussian Peaks} & \textbf{Starlet Peaks} &
\textbf{Multi-Gaussian Peaks}\\
\hline
41 bins   &$10^{5}$ &$10^{6}$ &$10^{7}$  \\
50 bins &$10^{5}$ & $10^{16}$ & -  \\
\hline
\end{tabular}
\caption{\label{tab:condition_number}Values of the Condition Number for the data covariance matrices. The smaller the number the easier it is to invert the matrix. In this case we get very large values for this estimator, leading to the conclusion that the data covariance matrices for Starlet Peaks and Gaussian Peaks - show very singular behaviour.}
\end{table*}
\subsubsection{Matrix condition number}\label{subsec:Matrix_condition_number}
To estimate how difficult it is to invert our data covariance matrices, we compute the corresponding \textit{condition number}: if the matrix is singular, the associated condition number is infinite, i.e. matrices with large condition numbers are more difficult to invert. We compute the condition number through the 2-norm of the matrix using singular value decomposition (SVD). As shown in \autoref{tab:condition_number}, the condition number depends on the binning choice, especially for the multi-scale analysis. Indeed, we find that choosing 41 linearly spaced bins for the peak counts instead of 50 reduces the condition number of the starlet peaks of about 10 orders of magnitude. For this reason we choose 41 bins when performing inference using peak counts.

\subsection{MCMC simulations and posterior distributions}
To explore and constrain the parameter space, we use the \url{emcee} package, which is a  python implementation of the affine-invariant ensemble sampler for Markov chain Monte Carlo (MCMC) introduced by \citep{2013PASP..125..306F}. The pipeline is built in a way that both the computation of the power spectrum and peak counts along with the MCMC are run in parallel to gain computation time. We assume a flat prior, specifically following \citep{Coulton_2019}, a Gaussian likelihood function as defined in Eq. \eqref{Likelihood_function}, and a model-independent covariance matrix as discussed in Sec. \ref{Section:covariance}. The walkers are initialised in a tiny Gaussian ball of radius $10^{-3}$ around the fiducial cosmology $[M_{\nu},\Omega_{\rm m},10^{9}A_{\rm s}]=[0.1,0.3,2.1]$ and we estimate the posterior using 120 walkers. Our chains are stable against the length of the chain, and we verify their convergence by employing \textit{Gelman Rubin} diagnostics \citep{gelman1992}. To plot the contours we use the \url{ChainConsumer} python package \citep{Hinton2016}.
\\

\section{Results}\label{subsec:results}

\begin{table*}
\begin{tabular}{lccccc}
\hline
\hline
\textbf{FoM} & $(M_{\nu}$, $\Omega_{\rm m}$)& ($M_{\nu}$, $A_{\rm s}$) & ($\Omega_{\rm m}$, $A_{\rm s}$) & ($M_{\nu}$, $\Omega_{\rm m}$, $A_{\rm s}$) \\
\hline
Power spectrum & 1585 & 77 & 1079 & 2063  \\
Single-Gaussian Peaks & 3559 & 200 & 2861 & 6537 \\
Single-Gaussian Peaks + PS & 5839 & 322 & 4688 & 11205 \\
Starlet Peaks & 7818 & 434 & 6428 & 12755 \\
Starlet Peaks diagonal & 8434 & 471 & 6936 & 16528  \\
Starlet Peaks + PS & 9796 & 540 & 7966 & 16166 \\
Multi-Gaussian Peaks & 11804 & 655 & 9729 & 18647 \\
Multi-Gaussian Peaks diagonal & 13612 & 770 & 11231 & 29780 \\
Multi-Gaussian Peaks + PS & 13471 & 742 & 10983 & 22115 \\
\hline
\end{tabular}
\caption{\label{tab: Table_FoM}Values of the FoM as defined in Eq. \eqref{eq:Figure_of_Merit} for the different parameters pairs ($\alpha$, $\beta$) for each observable employed in the likelihood analysis: the power spectrum, the peaks counted on maps smoothed with the kernel in consideration and Peaks + PS always refer to the constraints obtained with the peaks relative to some filters and the power spectrum while the term \textit{diagonal} refers to the contours obtained with a likelihood analysis performed by only considering the diagonal elements of the data covariance matrix. We provide in the last column the 3D FoM given as the inverse of the volume in ($M_{\nu}, A_{\rm s}, \Omega_{\rm m}$) space.} 
\end{table*}

\begin{table*}[!htbp]
\begin{tabular}{lcccccccc}
\hline
\hline

\textbf{FoC} & ($M_{\nu}$, $\Omega_{\rm m}$) & ($M_{\nu}$, $A_{\rm s}$) & ($\Omega_{\rm m}$, $A_{\rm s}$)   \\
\hline
Power spectrum & 1.00  &1.98 &1.10 \\
Single-Gaussian Peaks  &1.21 &1.42  &1.01 \\
Single-Gaussian Peaks + PS &1.19  &1.31  &1.04\\
Starlet Peaks  &1.14  &1.18  &1.14\\
Starlet Peaks + PS  &1.13  &1.13  &1.20\\
Multi-Gaussian Peaks  &1.13  &1.16  &1.17\\
Multi-Gaussian Peaks + PS &1.13  &1.14  &1.18\\
\hline
\end{tabular}
\caption{\label{Figure_of_correlation}Value of the Figure of Correlation for each pair of cosmological parameters corresponding to the different tomographic observables: the power spectrum alone (PS), the Peaks alone for different filters and the two statistics combined (Peaks + PS). As explained in the text, FoC = 1 corresponds to uncorrelated parameters, while the further the FoC is to 1, the more correlations are present. Qualitatively, this can be appreciated by looking at the inclination of the contours: by looking at Fig. \ref{fig:Single_scale_vs_Multi-scale} we can see more oblique contours for the Gaussian peaks in the plane ($M_{\nu}$, $\Omega_{\rm m}$) compared to the power spectrum, while for the pair ($M_{\nu}$, $A_{\rm s}$) the power spectrum shows more correlation than peaks.}
\end{table*}

We now illustrate forecast results on the sum of neutrino masses $M_{\nu}$, on the matter density parameter $\Omega_{\rm m}$ and on the power spectrum amplitude $A_{\rm s}$ for a survey with \Euclid - like noise in a tomographic setting with four source redshifts $z_s=[0.5,1.0,1.5,2.0]$, and compare results for different observables (power spectrum and peak counts) and filters (single-Gaussian,  starlet and multi-Gaussian).

\subsection{Single-scale vs multi-scale analysis}

\begin{figure*}[!htbp]
       \includegraphics[width=0.45\textwidth]{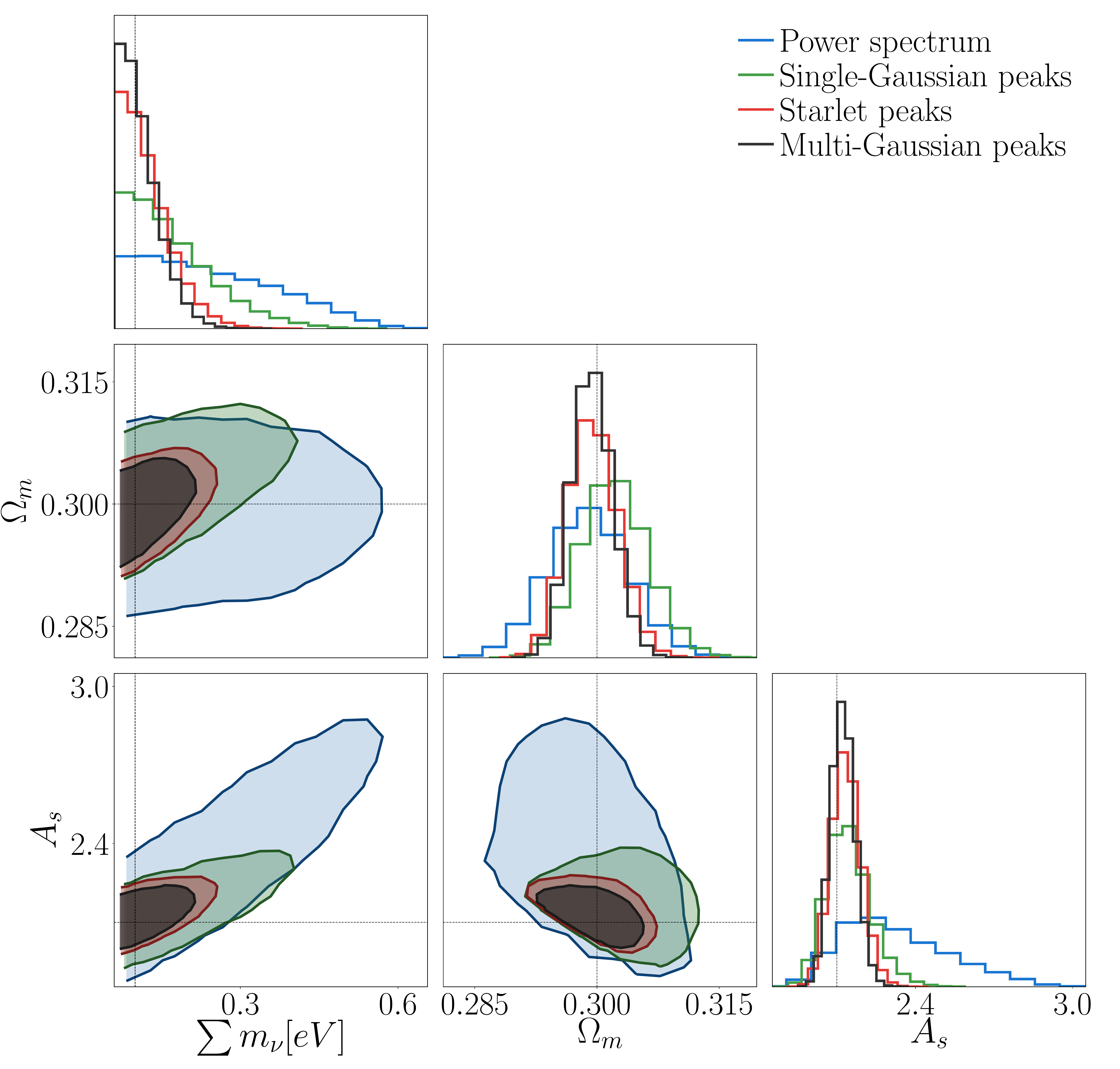} 
        \includegraphics[width=0.45\textwidth]{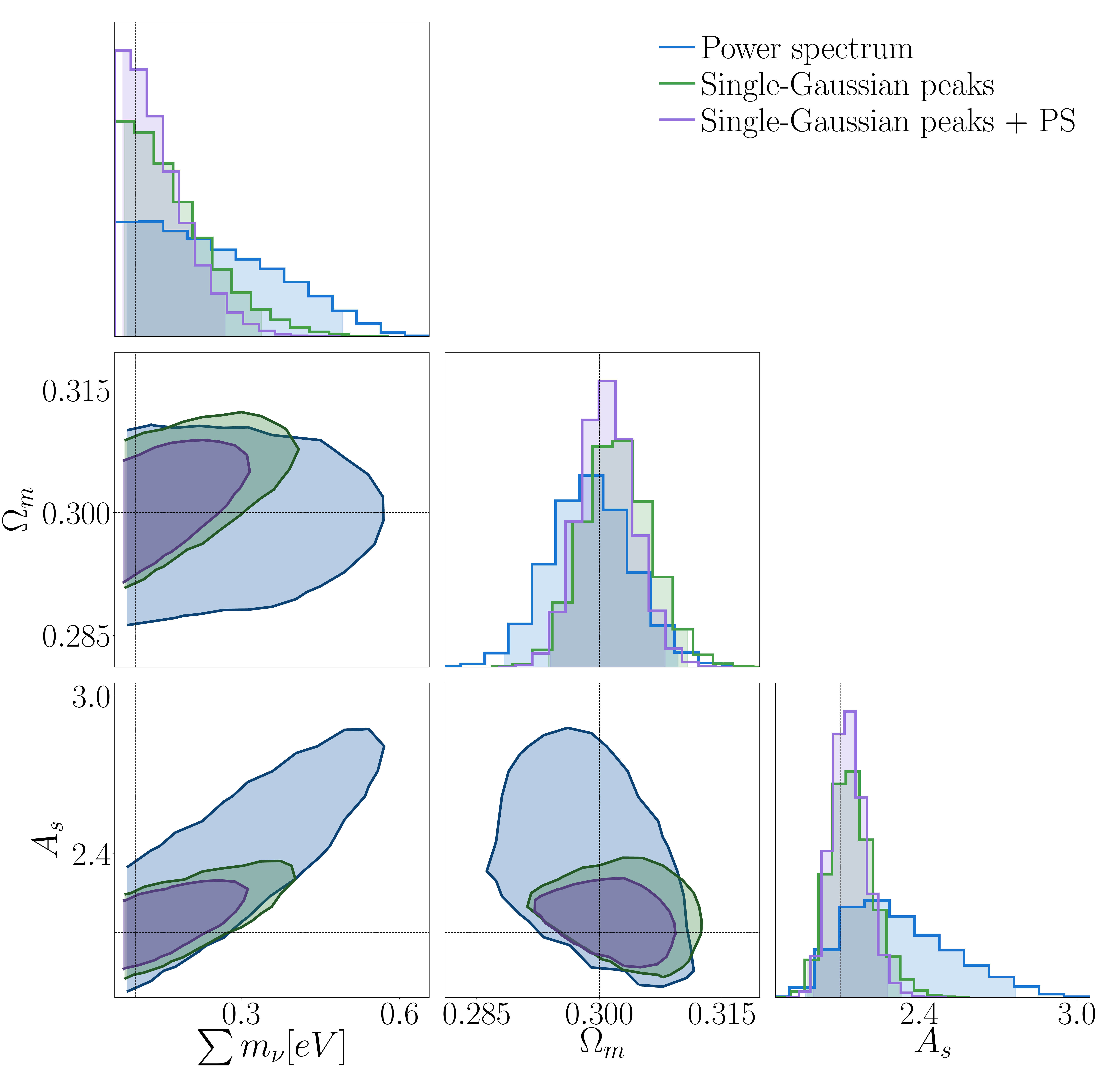}
  \caption{\label{fig:Single_scale_vs_Multi-scale}95 \%  confidence contours tomography with source redshifts $z_s=[0.5,1.0,1.5,2.0]$ and corresponding galaxy number density: $n_\mathrm{gal}=[11.02, 11.90, 5.45, 1.45]$. The black dotted line is the fiducial model: $[\sum m_{\nu}, \Omega_{\rm m}, 10^9A_{\rm s}] = [0.1, 0.3, 2.1]$. \textbf{Left panel}: constraints from power spectrum (blue contours) computed on noisy maps smoothed with a Gaussian kernel $\theta_\mathrm{ker}=1$ arcmin, constraints from Gaussian Peak counts (green contours) computed on noisy maps smoothed with a Gaussian kernel $\theta_\mathrm{ker}=1.6$ arcmin, constraints from Starlet Peak counts (red contours) computed on noisy maps smoothed with a Starlet kernel with corresponding resolutions [1.6, 3.2, 6.4] arcmin + coarse map, constraints from multi-Gaussian Peak counts (black contours) computed on noisy maps smoothed with a multi-Gaussian kernel with corresponding resolutions [1.2, 2.7, 5.5, 9.5] arcmin. \textbf{Right panel}: constraints from power spectrum (blue contours) computed on noisy maps smoothed with a Gaussian kernel $\theta_\mathrm{ker}=1$ arcmin, constraints from Gaussian Peak counts (green contours) computed on noisy maps smoothed with a Gaussian kernel $\theta_\mathrm{ker}=1.6$ arcmin and the two statistics jointly (violet contours).}
\end{figure*}

In the left panel of Fig. \ref{fig:Single_scale_vs_Multi-scale} we compare constraints obtained from the single-scale and the multi-scale peak counts analysis against the power spectrum contours. For the single scale we employ a Gaussian filter with $\theta_{ker}^{G} = 1.6'$. For the multi-scale analysis we employ a starlet filter and a concatenation of Gaussian filters with smoothing widths chosen such that the profiles match the starlet scales, as described in Sec. \ref{subsec:filters}. More specifically, we show the comparison among the power spectrum (blue contours), the single-scale peaks (green contours), the starlet peaks (red contours) and the multi-Gaussian peaks (black contours).  We confirm that peak counts outperform power spectrum constraints even in the single-Gaussian case, as found in \citep{PhysRevD.99.063527}. In addition, we find that a multi-scale approach leads to a remarkable improvement with respect to a single-scale approach in terms of constraining power, as expected, due to its higher information content concerning structure formation. 

We quantify these outcomes by considering the Figure of Merit defined in Eq. \eqref{eq:Figure_of_Merit}. As shown in \autoref{tab: Table_FoM}, the FoM for the single-Gaussian peaks in the parameter space plane ($M_{\nu}$, $\Omega_{\rm m}$) is more than twice that given by the power spectrum, the one from starlet peaks is more than twice that obtained with the single-Gaussian peaks, and the multi-Gaussian peaks FoM is more than three times that of the single-Gaussian case. Concerning the ($M_{\nu}$, $A_{\rm s}$) and ($\Omega_{\rm m}$, $A_{\rm s}$) planes, the FoM for Gaussian peaks is about three times that for the power spectrum, the one from starlet peaks is again about twice that obtained with the single-Gaussian peaks, and the multi-Gaussian peaks give again a FoM about three times that seen for single-Gaussian peaks contours.

As further investigation, we compute the Figure of Correlation as defined in Eq. \eqref{Eq:fig_of_correlation} to study the correlation among the parameters. By looking at \autoref{Figure_of_correlation}, one can see how values for the power spectrum for the pairs ($M_{\nu}$, $\Omega_{\rm m}$) and ($\Omega_{\rm m}$, $A_{\rm s}$) are close to one, suggesting that correlation among them appears to be very small, while the plane ($M_{\nu}$, $A_{\rm s}$) shows more correlation, as its FoC is nearly twice as large. Qualitatively this can be appreciated by looking at the inclination of the contours. More specifically, concerning the plane ($M_{\nu}$, $\Omega_{\rm m}$), the power spectrum contours are horizontal and show also visually that these two parameters are not correlated; constraints obtained via peak counts show a slightly larger correlation, increasing by 21\% for Gaussian peaks and by about 13-14\% for the multi-scale analysis, with respect to the power spectrum. It is interesting to note that in the ($M_{\nu}$, $A_{\rm s}$) plane, the correlation decreases by 30\% when using single-Gaussian peaks and by about 40\% when using a multi-scale approach compared to the power spectrum, suggesting that peak counts can play an important role in breaking the degeneracy for this pair of parameters. Independently of the correlation, all constraints obtained with multi-scale filtering are tighter than the ones obtained via single-scale filtering, and both are tighter than the ones for the power spectrum.

\subsection{Joint contours}
Based on the previous result, we are now interested in the constraints obtained when considering the two statistics jointly and on the impact of the different filter settings in this context. In particular, by focusing on the right panel of Fig. \ref{fig:Single_scale_vs_Multi-scale}, where we compare the 95\% confidence contours of the power spectrum (blue contours) with the single-Gaussian peaks (green contours) and the two joint statistics (violet contours), we notice how in the single-scale approach the addition of the power spectrum to the peak counts brings a non-negligible improvement in terms of constraining power with respect to the peaks alone. More specifically, reading the values presented in \autoref{tab: Table_FoM}, we see that the FoM of the joint contours is roughly $1.6$ times that of the peaks alone, and more the three times that of the power spectrum.

\begin{figure*}[!htbp]
       \includegraphics[width=0.45\textwidth]{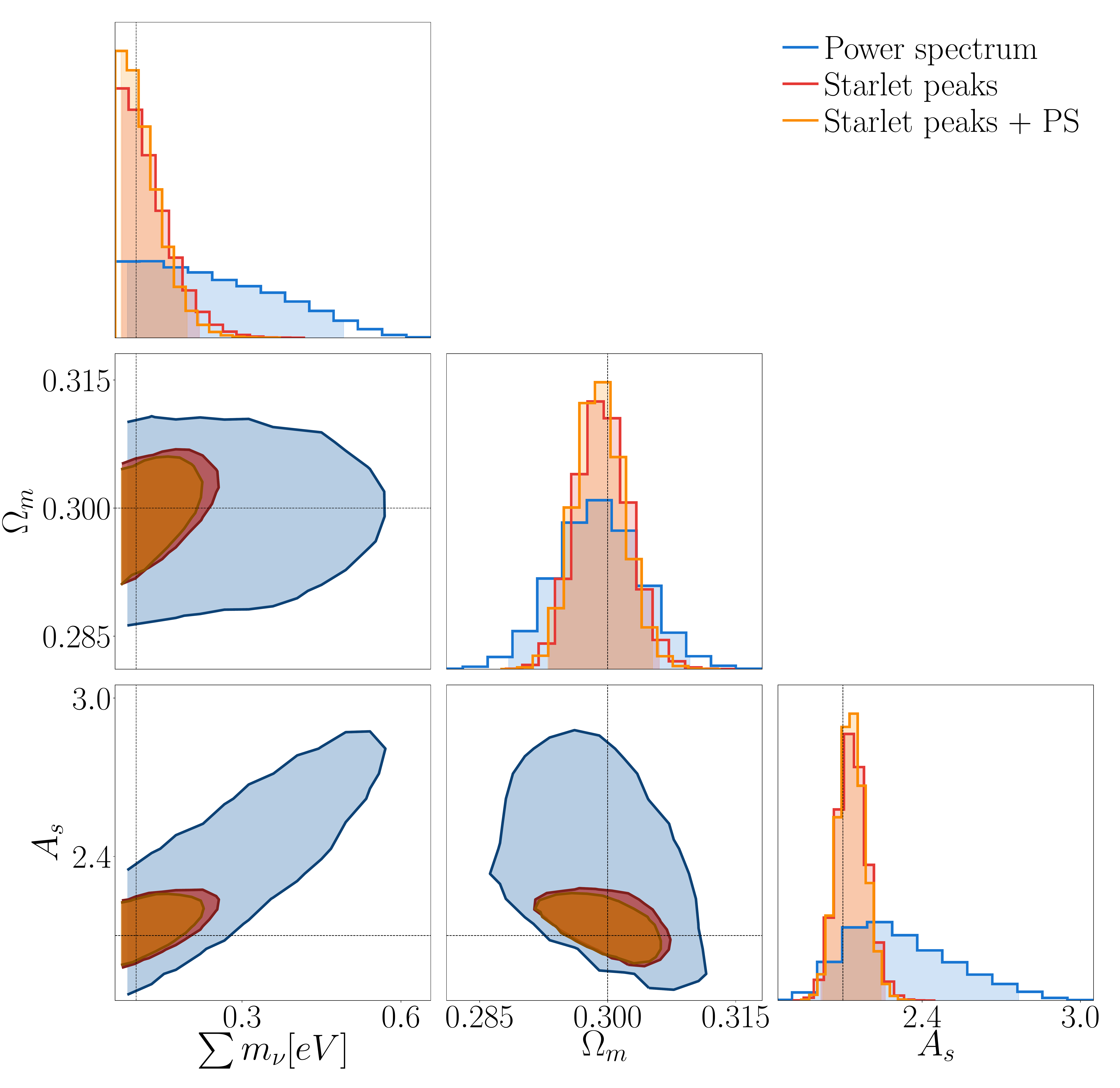} 
        \includegraphics[width=0.45\textwidth]{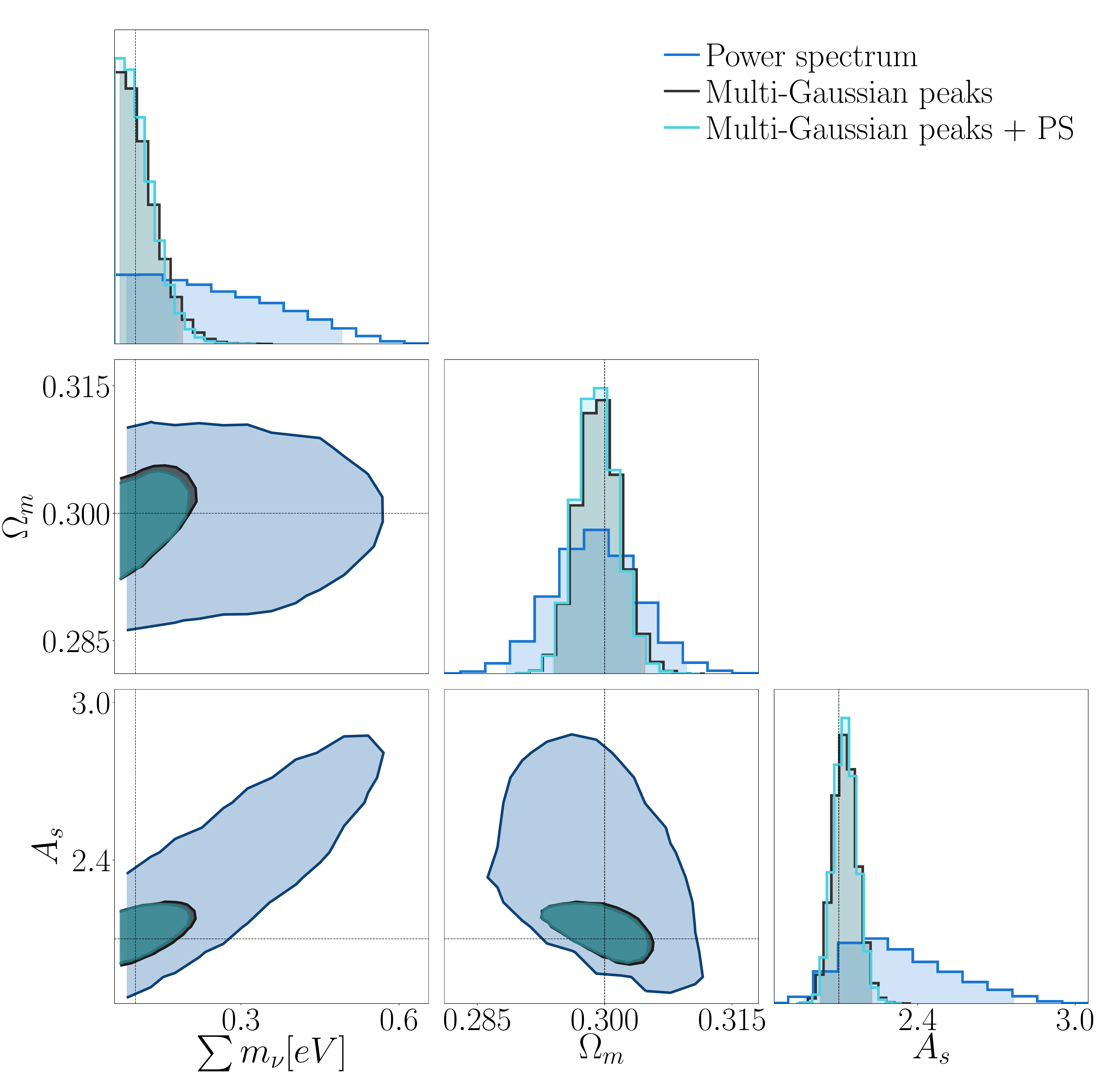}
  \caption{\label{fig:multi_scale_joints}95 \%  confidence contours tomography with source redshifts $z_s=[0.5,1.0,1.5,2.0]$ and corresponding galaxy number density: $n_\mathrm{gal}=[11.02, 11.90, 5.45, 1.45]$. The black dotted line is the fiducial model: $[\sum m_{\nu}, \Omega_{\rm m}, 10^9A_{\rm s}] = [0.1, 0.3, 2.1]$. \textbf{Left panel}: constraints from power spectrum (blue contours) computed on noisy maps smoothed with a Gaussian kernel $\theta_\mathrm{ker}=1$ arcmin, constraints from Starlet Peak counts (red contours) computed on noisy maps smoothed with a Starlet kernel with corresponding resolutions [1.6, 3.2, 6.4] arcmin + coarse map and constraints from the two statistics joint (orange contours). \textbf{Right panel}: constraints from power spectrum (blue contours) computed on noisy maps smoothed with a Gaussian kernel $\theta_\mathrm{ker}=1$ arcmin, constraints from multi-Gaussian Peak counts (black contours) computed on noisy maps smoothed with a  multi-Gaussian kernel with corresponding resolutions [1.2, 2.7, 5.5, 9.5] arcmin and the two statistics jointly (light blue contours).}
\end{figure*}

\begin{figure*}
    \centering
    \includegraphics[width=1.\textwidth]{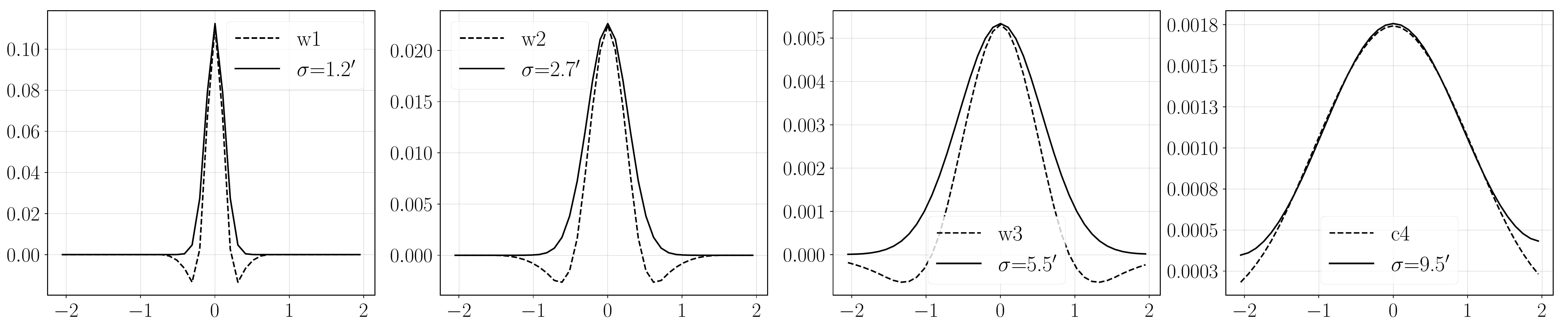}
    \caption{We show the matching between the Gaussian filter and the starlet at the different scales. We chose the kernel for the multi-Gaussian concatenation such that the maximum of the two profiles match. From left to right are the finest scale to the smoothest scale where $[w1, w2, w3, c4]=[1.6, 3.2, 6.2, 12.8]$ arcmin.}
    \label{fig:gaussian_starlet_functions}
\end{figure*}

Focusing now on the multi-scale approaches, in Fig. \ref{fig:multi_scale_joints} we show the same comparison in the left panel by comparing the power spectrum with the starlet peaks alone (red contour) and the two joint statistics (orange contours) and in the right panel for the multi-Gaussian case  with the multi-Gaussian peaks alone (black contours) and the two joint statistics (turquoise contours). In Fig. \ref{fig:gaussian_starlet_functions} we show the matching between the Gaussian filter and the starlet at the different scales to show how we chose the kernel for the multi Gaussian concatenation (such that the maximum of the two profiles matches). We notice in this case that the FoM of the combined statistics are roughly $1.1-1.2$ times the peaks alone in the starlet case and $1.1$  times the peaks alone in the multi-Gaussian case, suggesting that the information given by the joint statistics is mostly contained in the multi-scale peak counts alone. Peak counts therefore appear to be competitive and sufficient statistics for parameter inference when dealing with weak lensing convergence maps as input data. This further confirms that lensing peaks are a powerful tool in the context of cosmological parameter inference, emphasizing as well the importance of the role played by the filtering choice.

\subsection{Marginalised constraints}
\begin{figure*}[h]
    \includegraphics[width=\textwidth]{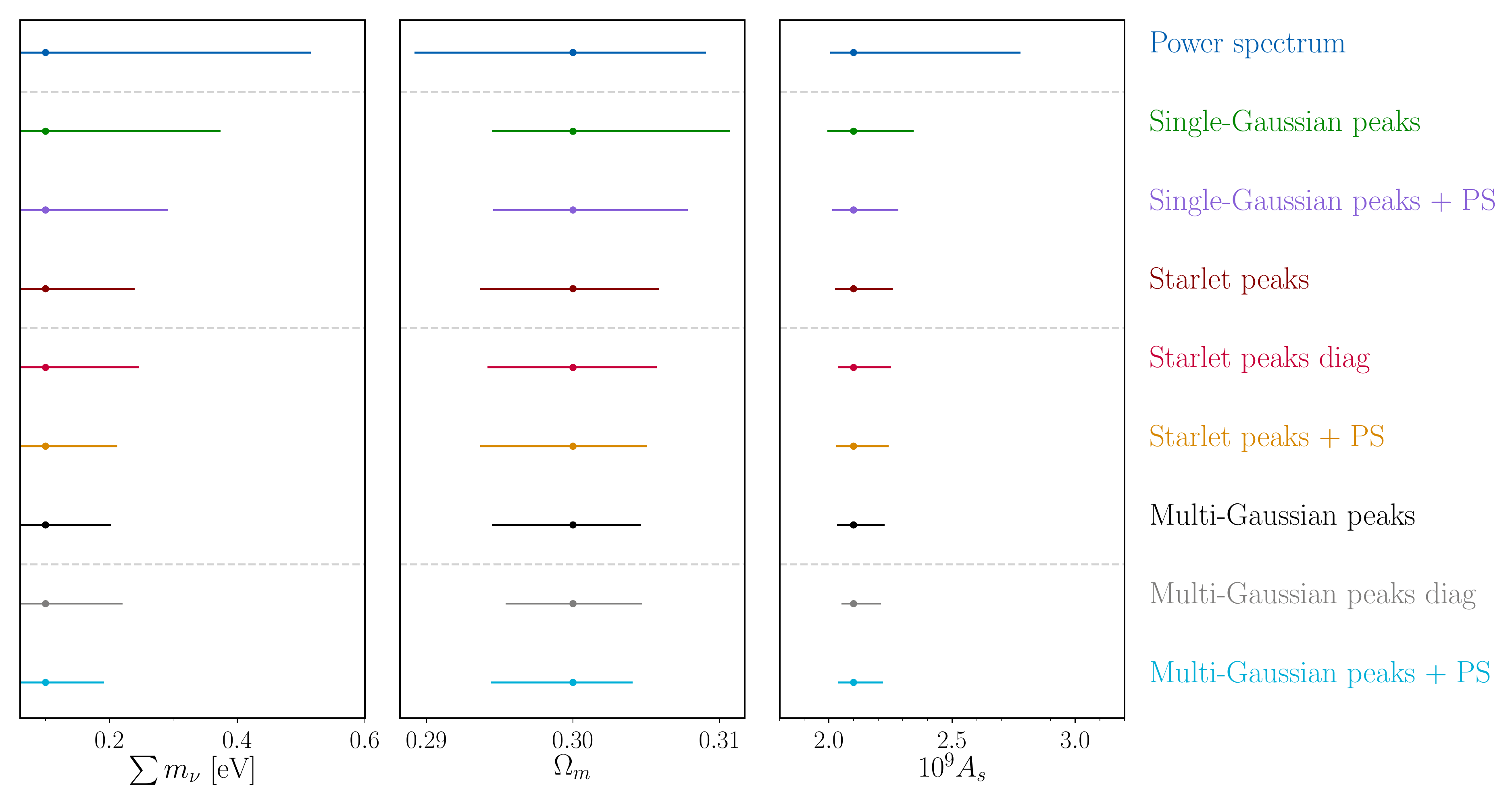}
    \caption{\label{fig:Marginalised_plot}Marginalized constraints on each parameter for forecasts showing the 2.5 and 97.5 percentiles with respect to the fiducial model. These marginalised constraints refer to a tomographic setting with $z=[0.5, 1.0, 1.5, 2.0]$ with the fiducial model set at $ [M_{\nu}, \Omega_{\rm m}, 10^{9}A_{\rm s}]=[ 0.1, 0.3, 2.1 ]$ corresponding to the different observables employed within the likelihood analysis. The values are listed in Table \ref{tab:marginalised_parameters}.}
\end{figure*}

In Fig. \ref{fig:Marginalised_plot} we show the marginalised constraints on each cosmological parameter corresponding to the different observables. To compare the improvement obtained by employing the different statistics we compute the 1$\sigma$ marginalised error for each parameter, summarised in \autoref{tab:marginalised_errors}. In particular, we find an improvement of 35\%, 20\% and 58\% respectively on $M_{\nu}$, $\Omega_{\rm m}$ and $A_{\rm s}$ when employing the single-Gaussian peaks instead of the power spectrum, an improvement of 63\%, 40\% and 72\% when employing the starlet peaks instead of the power spectrum alone, and an improvement of 70\%, 40\% and 77\% when employing the multi-Gaussian peaks instead of the power spectrum alone. Namely, the starlet peaks outperform the single-Gaussian peaks by 43\% on $M_{\nu}$, 25\% on $\Omega_{\rm m}$ and 34\% on $A_{\rm s}$, and the multi-Gaussian peaks outperform the single-Gaussian peaks by 54\% on $M_{\nu}$, 25\% on $\Omega_{\rm m}$ and 45\% on $A_{\rm s}$. Finally, employing a multi-Gaussian instead of a starlet filter in the context of peak counts might improve the constraints by 19\% on $M_{\nu}$, and 18\% on $A_{\rm s}$, while no improvement is noticed for $\Omega_{\rm m}$.

\subsection{Starlet scales impact }
The left panel of Fig. \ref{fig:starlet_scales_contours} shows the impact of the different starlet decomposition scales on the constraints. The MCMC chain used for the starlet decomposition results of the analysis has been obtained by considering all starlet scales, i.e. $[1.6, 3.2, 6.4]$ arcmin $+$ coarse scale, shown in red. To check that we are allowed to exclude the finest scale in the entire analysis, namely not to include the resolution corresponding to $0.8$ arcmin - which won't satisfy the survey requirements - we compare the constraints relative to $[0.8, 1.6, 3.2, 6.4]$ arcmin $+$ coarse scale with the ones for $[1.6, 3.2, 6.4]$ arcmin $+$ coarse scale and we verify that they overlap. We then investigate the impact of the different starlet scales and we obtain that it is sufficient to consider the setting $[3.2, 6.4]$ $+$ coarse scale to obtain results competitive with the full set of scales, as shown by the dark blue contours in the figure that match the full starlet decomposition contours. Hence, we identify $w_2=3.2$ arcmin as the smallest scale needed to obtain the maximal constraints with convergence maps of resolution $0.4$ arcmin. We also perform the inference by adding one scale at a time in the observable array to show how the contours shrink as a function of the number of starlet scales. We find that the only setting that recovers almost the full information is given by [3.2, 6.4, coarse]. We also notice from the contours relative to $[w_1, w_2, w_3]$ that excluding the coarse scale leads to a loss of information (precisely 28\% on $M_{\nu}$, 33\% on $\Omega_{\rm m}$ and 19\% on $A_{\rm s}$).


\begin{table*}
\begin{tabular}{lccccc}
\hline
\hline
\textbf{Observable} & $\mathbf{M_{\nu}+}$ & $\mathbf{\Omega_{\rm m}-}$ & $\mathbf{\Omega_{\rm m}+}$ & $\mathbf{A_{\rm s}-}$ & $\mathbf{A_{\rm s}+}$  
\\
\hline
Power spectrum & 0.514 & 0.289 & 0.308 & 2.008 & 2.774 \\
Single-Gaussian Peaks & 0.372 & 0.295 & 0.311 & 1.998 & 2.340  \\
Single-Gaussian Peaks + PS & 0.290 & 0.294 & 0.307 & 2.016 & 2.277  \\
Starlet Peaks & 0.238 & 0.293 & 0.305 & 2.029 & 2.255  \\
Starlet Peaks diagonal & 0.244 & 0.294 & 0.305 & 2.039 & 2.248 \\
Starlet Peaks + PS & 0.210 & 0.293 & 0.305 & 2.032 & 2.239 \\
Multi-Gaussian Peaks  & 0.201 & 0.294 & 0.305 & 2.037 & 2.222  \\
Multi-Gaussian Peaks diagonal  & 0.218 & 0.295 & 0.305 & 2.054 & 2.207 \\
Multi-Gaussian Peaks + PS & 0.190 & 0.294 & 0.304 & 2.042 & 2.216 \\
\hline
\end{tabular}
\caption{\label{tab:marginalised_parameters}Values of the 2.5 and 97.5 percentiles for each cosmological parameter as illustrated in Fig. \ref{fig:Marginalised_plot}. In this table we also show the values corresponding to the marginalised constraints obtained using only the diagonal elements of the covariance matrices. They are very similar to the ones obtained by employing the full covariance. We further investigate this aspect in Sec. \ref{subsec:covariances}.}
\end{table*}

\begin{table}
\begin{tabular}{lccc}
\hline
\hline
$\mathbf{\sigma_{\alpha\alpha}}$ & $\mathbf{M_{\nu}}$ & $\mathbf{\Omega_{\rm m}}$ & $\mathbf{A_{\rm s}}$ \\
\hline
Power spectrum & 0.127 & 0.005 & 0.204\\
Single-Gaussian Peaks & 0.083 & 0.004 & 0.086\\
Single-Gaussian Peaks + PS & 0.061 & 0.003 & 0.066\\
Starlet Peaks & 0.047 & 0.003 & 0.057\\
Starlet Peaks diagonal & 0.049 & 0.003 & 0.053\\
Starlet Peaks + PS & 0.040 & 0.003 & 0.052\\
Multi-Gaussian Peaks & 0.038 & 0.003 & 0.047\\
Multi-Gaussian Peaks diagonal & 0.042 & 0.002 & 0.039\\
Multi-Gaussian Peaks + PS & 0.035 & 0.002 & 0.044\\

\hline
\end{tabular}
\caption{\label{tab:marginalised_errors}Values of 1-$\sigma$ marginalised error for each cosmological parameter for the different observables.}
\end{table}

\subsection{Covariances: Gaussian multi-scale and Starlet comparison}\label{subsec:covariances}

\begin{figure*}[!htbp]
       \includegraphics[width=0.45\textwidth]{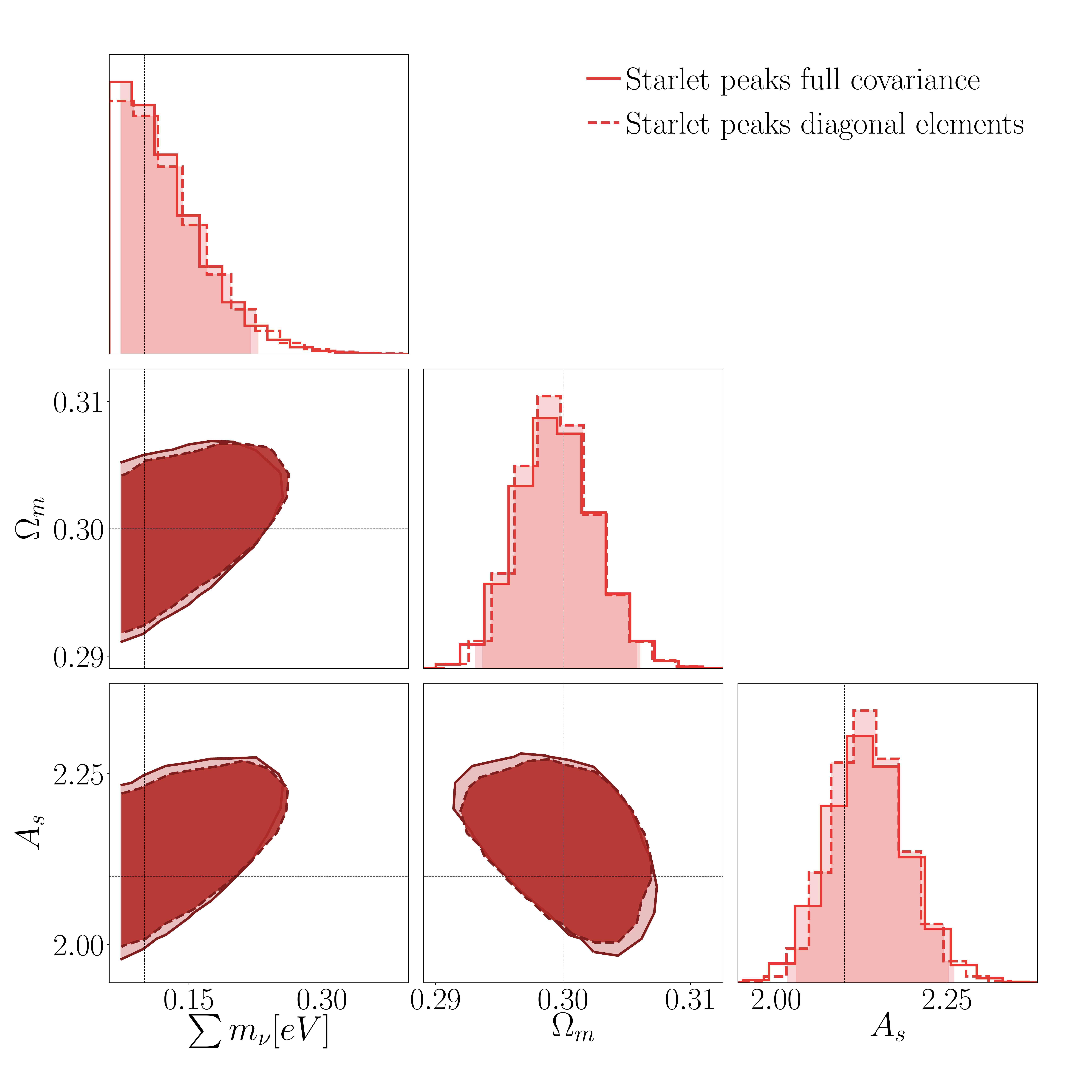} 
        \includegraphics[width=0.45\textwidth]{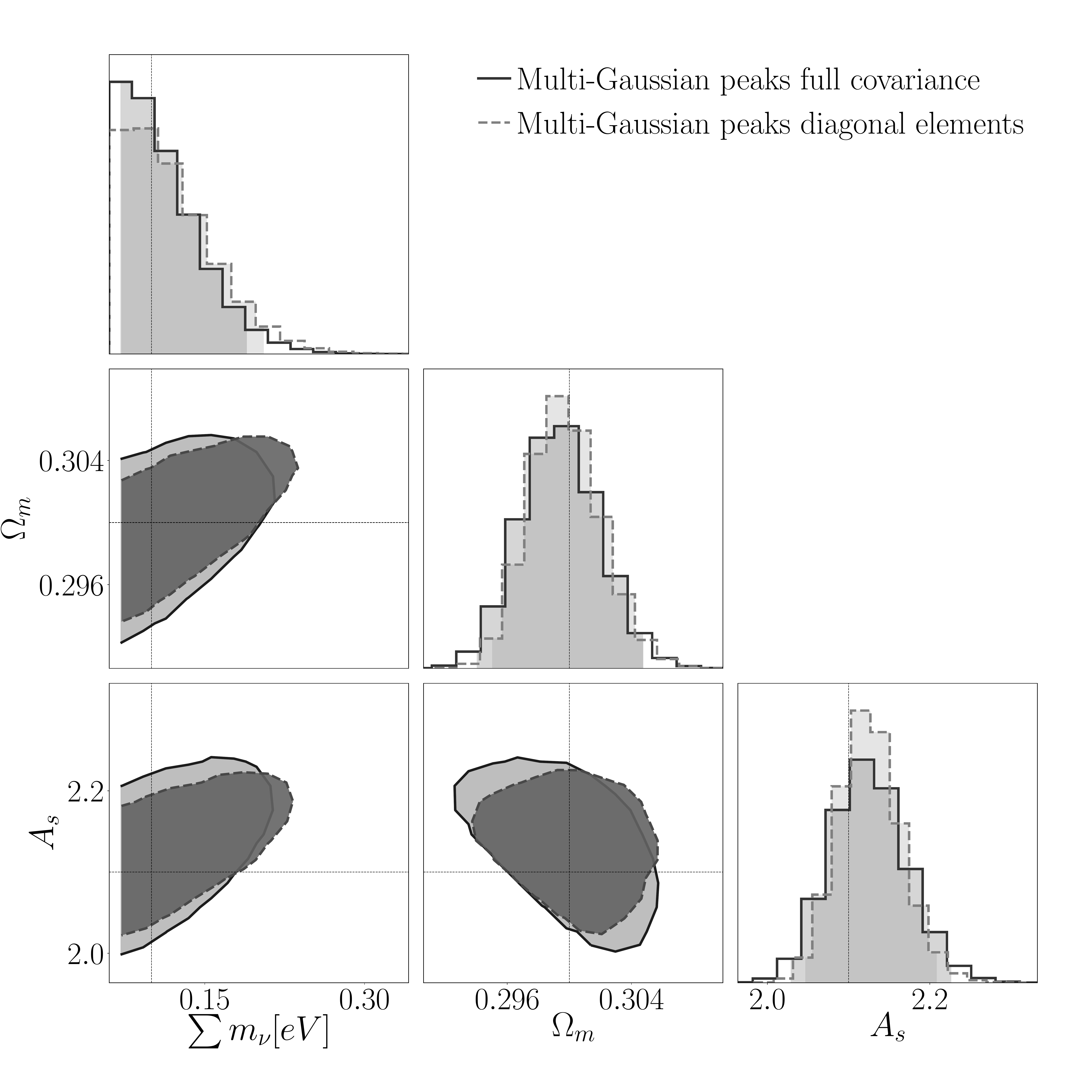}
  \caption{\label{fig:full_cov_vs_diagonal}95 \%  confidence contours tomography with redshifts $z_s=[0.5,1.0,1.5,2.0]$ and corresponding galaxy number density: $n_\mathrm{gal}=[11.02, 11.90, 5.45, 1.45]$. The black dotted line is the fiducial model: $[\sum m_{\nu}, \Omega_{\rm m}, 10^9A_{\rm s}] = [0.1, 0.3, 2.1]$. \textbf{Left panel}: constraints from starlet peak counts (continuous red contours) obtained employing the full covariance matrix against constraints from starlet peak counts (dashed red contours) obtained employing the diagonal elements only of the covariance matrix in the likelihood analysis. \textbf{Right panel}: constraints from multi-Gaussian peak counts (continuous black contours) obtained employing the full covariance matrix against constraints from multi-Gaussian peak counts (dashed gray contours) obtained employing the diagonal elements only of the covariance matrix in the likelihood analysis}
\end{figure*}

In this section we investigate the impact of the choice of the filter on the data covariance matrix. Indeed, by looking at the correlation matrices of Fig. \ref{fig:Covariances} it is clear that the starlet (left panel) has the tendency to \textit{diagonalize} the matrix while the multi-Gaussian case (right panel) presents non-trivial off-diagonal terms as introduced in Sec. \ref{Section:covariance}. To further explore this aspect we have run the likelihood analysis considering just the diagonal elements of the covariance matrices in order to compare results with full covariance case. We find the constraints illustrated in Fig. \ref{fig:full_cov_vs_diagonal}: in the left panel we plot the starlet peak counts contours obtained with the full covariance matrix (red) against the diagonal-only version of the data covariance matrix (dashed dark red). On the right panel we show the same comparison for the multi-Gaussian peaks case with the full covariance case (black) against diagonal-only contours (dashed gray). We see that for the starlet filter the majority of the information is indeed encoded in the diagonal elements, while for the multi-Gaussian case the presence of non-trivial correlations among the scales makes the contours slightly larger for $\Omega_{\rm m}$ and $A_{\rm s}$, while it adds some information on $M_{\nu}$ with respect to the diagonal case. We can quantify this by taking the ratio between the FoM relative to the full covariance and the diagonal elements cases: for the starlet we find a ratio of $1.07$ and for the multi-Gaussian $1.15$. The same interpretation arises when looking at the 1-$\sigma$ marginalised error of \autoref{tab:marginalised_errors}: excluding the off-diagonal terms in the starlet data covariance matrix implies a loss of information of $4\%$ on $M_{\nu}$, no loss on $\Omega_{\rm m}$ and a gain of $7\%$ on $A_{\rm s}$. For the multi-Gaussian case, the same procedure leads to a loss of $11\%$ on $M_{\nu}$, a gain of $33\%$ on $\Omega_{\rm m}$ and a gain of $17\%$ on $A_{\rm s}$. This is an interesting aspect of the starlet filter that could prove to be useful when dealing with high dimensional data and the covariance matrix can be difficult to invert.

\section{Conclusions}\label{subsec:Conclusions}
In this paper, we infer the sum of neutrino masses $M_{\nu}$, the matter density parameter $\Omega_{\rm m}$ and the amplitude of the primordial power spectrum $A_{\rm s}$ for a survey with \Euclid-like noise using tomographic weak lensing. Our goal is to compare the constraining power of multi-scale filtering approaches, namely the starlet filter and a concatenation of Gaussian filters, with respect to a single-Gaussian one in the context of peak counts. We also compute the constraints with standard second-order statistics, in particular using the lensing power spectrum as a benchmark for the comparison. We compare the outcomes obtained from filtering the lensing convergence maps, which have a resolution of 0.4 arcmin, with a Gaussian kernel of smoothing size 1.6 arcmin, a starlet kernel, and a concatenation of Gaussians.
More specifically, the starlet filter is an isotropic undecimated wavelet transform that allows us to extract the information encoded in different spatial scales simultaneously. Setting the number of scales in the transform to four, the starlet kernel sizes for our maps correspond to [0.8, 1.6, 3.2, 6.4] arcmin + the coarse map, since the starlet transform returns maps filtered at dyadic scales. In deriving parameter constraints we exclude the first scale and work with [1.6, 3.2, 6.4] arcmin + coarse map. To fairly compare it with a multi-Gaussian, we set the standard deviations of the Gaussian kernels at each scale such that their profile peaks match the corresponding starlet scale peaks, resulting in a concatenation of Gaussians with standard deviations of [1.2, 2.7, 5.5, 9.5] arcmin respectively.

We find the following results:
\begin{itemize}
    \item[a)]{For peak counts, a multi-scale filtering approach of the noisy maps leads to an improvement factor of more than two over a single-scale approach (single-Gaussian kernel) for the joint constraints on $(M_{\nu}, \Omega_{\rm m})$, $(M_{\nu}, A_{\rm s})$ and $(\Omega_{\rm m}, A_{\rm s})$ when using a starlet kernel, and a factor of more than three when using a multi-Gaussian filter. This is even more evident in the marginalised constraints, where the improvement is respectively 43\% on $M_{\nu}$, 25\% on $\Omega_{\rm m}$ and 34\% on $A_{\rm s}$ for the starlet, while for the multi-Gaussian it is 54\% on $M_{\nu}$, 25\% on $\Omega_{\rm m}$ and 45\% on $A_{\rm s}$. Employing a multi-Gaussian instead of a starlet filter in the context of peak counts might improve the constraints by 19\% on $M_{\nu}$, and 18\% on $A_{\rm s}$, while no improvement is noticed for $\Omega_{\rm m}$. Finally, multi-scale peak counts in both perform better than the power spectrum on the set of parameters $\left\lbrace M_{\nu}, \Omega_{\rm m}, A_{\rm s} \right\rbrace$ respectively by 63$\%$, 40$\%$ and 72$\%$ when using a starlet filter and by 70$\%$, 40$\%$ and 77$\%$ when using a multi-scale Gaussian filter}.
    \item[b)]{When combining multi-scale
    peaks and the power spectrum, i.e. using a concatenation of peak counts and the power spectrum as the observed data vector, we find that the information is mostly encoded in the peaks alone (for certain parameters, such as $\Omega_{\rm m}$ in the starlet case, it is \textit{completely} encoded). This suggests that when adopting a multi-scale approach, it might be sufficient to work with the peaks alone}.
     \item[c)]{The inclusion of the coarse map when counting peaks preserves crucial information. Moreover, for maps with a pixel size of 0.4 arcmin, there exists a minimum resolution (i.e. smallest scale needed) for the starlet scales corresponding to $\theta_\mathrm{ker}=3.2$ arcmin to achieve maximal constraining power. This enables us to exclude the first two finest scales of the starlet decomposition, which correspond to the highest frequencies and are the most prone to the impact of noise, allowing for a faster and more efficient analysis.}
     \item[d)]{We notice that employing a starlet filter leads to a highly diagonal data covariance matrix, while for the multi-Gaussian filter the off-diagonal terms are prominent, and correlations among the different scales are non-negligible. In other words, the majority of the information in the starlet filter case is encoded in the diagonal elements of the covariance matrix. This is an interesting aspect of  the  starlet  filter  that  could  prove useful  when dealing  with  high  dimensional  data  where the  covariance matrix can be difficult to invert.}
\end{itemize}

\noindent In summary, we confirm that weak-lensing peak counts are a powerful tool to infer cosmological parameters, especially when investigating the non-linear regime where the impact of parameters such as the neutrino masses becomes relevant. We also point out the importance of adopting a multi-scale approach in the context of weak-lensing peak counts, which bring the advantage of analysing the information encoded at different scales simultaneously, thereby leading to tighter constraints than single-scale analysis. As we have shown in Fig. \ref{fig:Marginalised_plot}, the two multi-scale filters we have studied (the multi-Gaussian filter and the starlet filter), have similar constraints. This is expected, as we choose the Gaussian kernels such that each profile peak matches with a starlet scale. Minimal residual differences between the two filters may be related to the binning: while this is the same for both, it might be that the two filters are optimal with different choices of the binning. We leave the investigation of the optimal binning for both multiscale Gaussian and starlet peaks to future work. There is however an advantage, in using the starlet filter over a multi-Gaussian filter: the starlet has the tendency to remove the off-diagonal terms in the covariance matrix, hence making the matrix more diagonal, easy and faster to invert. Moreover, \citep{2012MNRAS.423.3405L} have proved that it offers a clear and significant time advantage over standard aperture mass algorithms for all scales of interest. We implemented a pipeline that allows us to go from simulated lensing convergence maps as input data to constraints on cosmological parameters as final output, employing different filtering techniques with second-order (the power spectrum) and higher-order statistics (peak-counts). Hence, a future project will be to generalise the pipeline in terms of flexibility of the input data, including systematic effects and modelling of the noise. In particular, being able to control systematic errors and baryonic effects is as important as the statistical power to guarantee a robust analysis. In the context of weak-lensing peak counts, baryons can change the shape of the distribution of peaks by increasing the low $S/N$ end and decreasing the high $S/N$ values by a few percent, as quantified by \cite{2019MNRAS.488.3340F}. It has been shown as well by \citep{2020MNRAS.495.2531C} that ignoring baryonic effects can lead to strong biases in inferences from peak counts and that in principle these biases can be mitigated without significantly degrading cosmological constraints when baryonic effects are modeled and marginalized. Concerning intrinsic alignment and noise uncertainty, we will make further investigations in future studies with the aim of including such modelling in our pipeline and to ultimately apply our pipeline to real data coming from future galaxy surveys.

\begin{figure*}
    \includegraphics[width=\columnwidth]{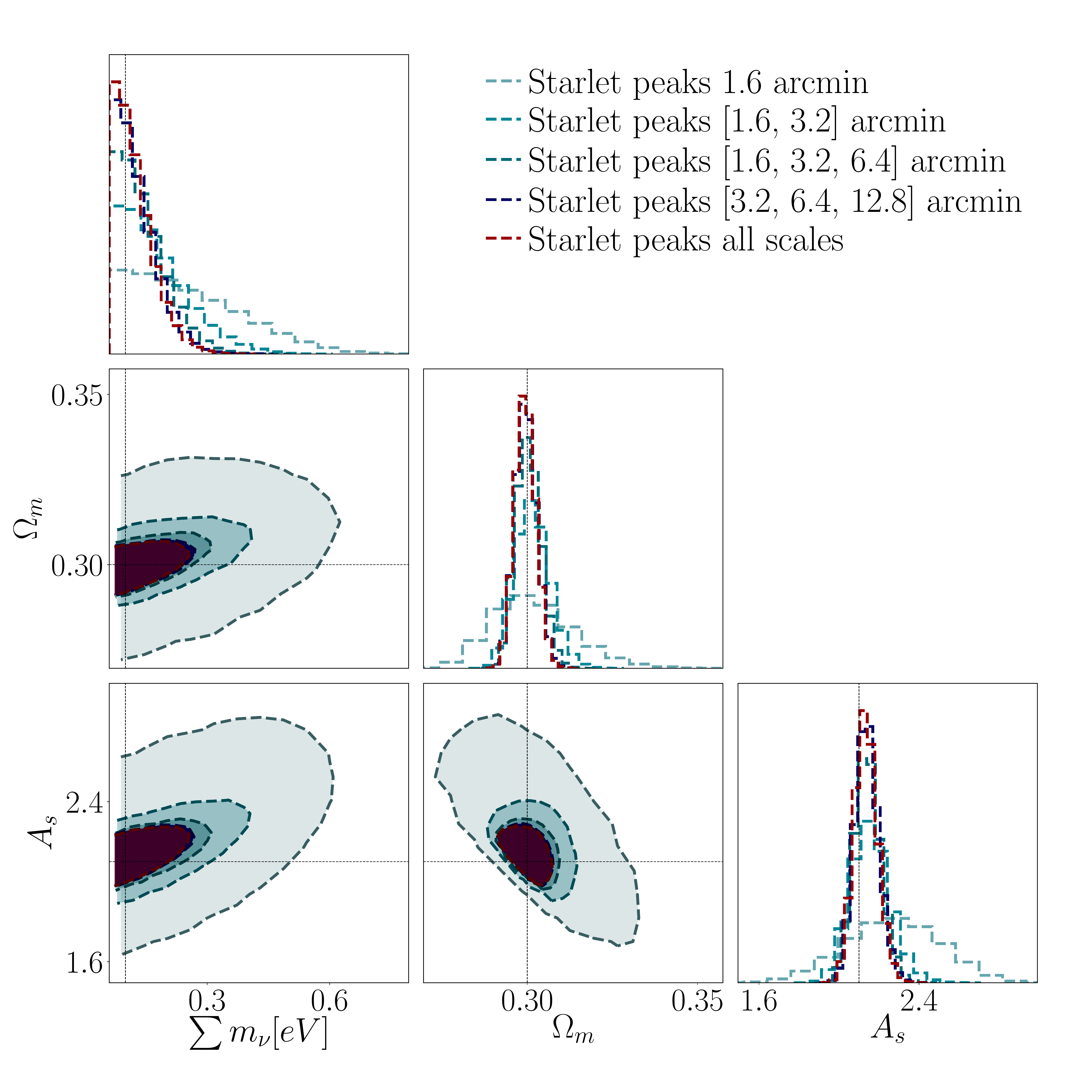}
        \includegraphics[width=\columnwidth]{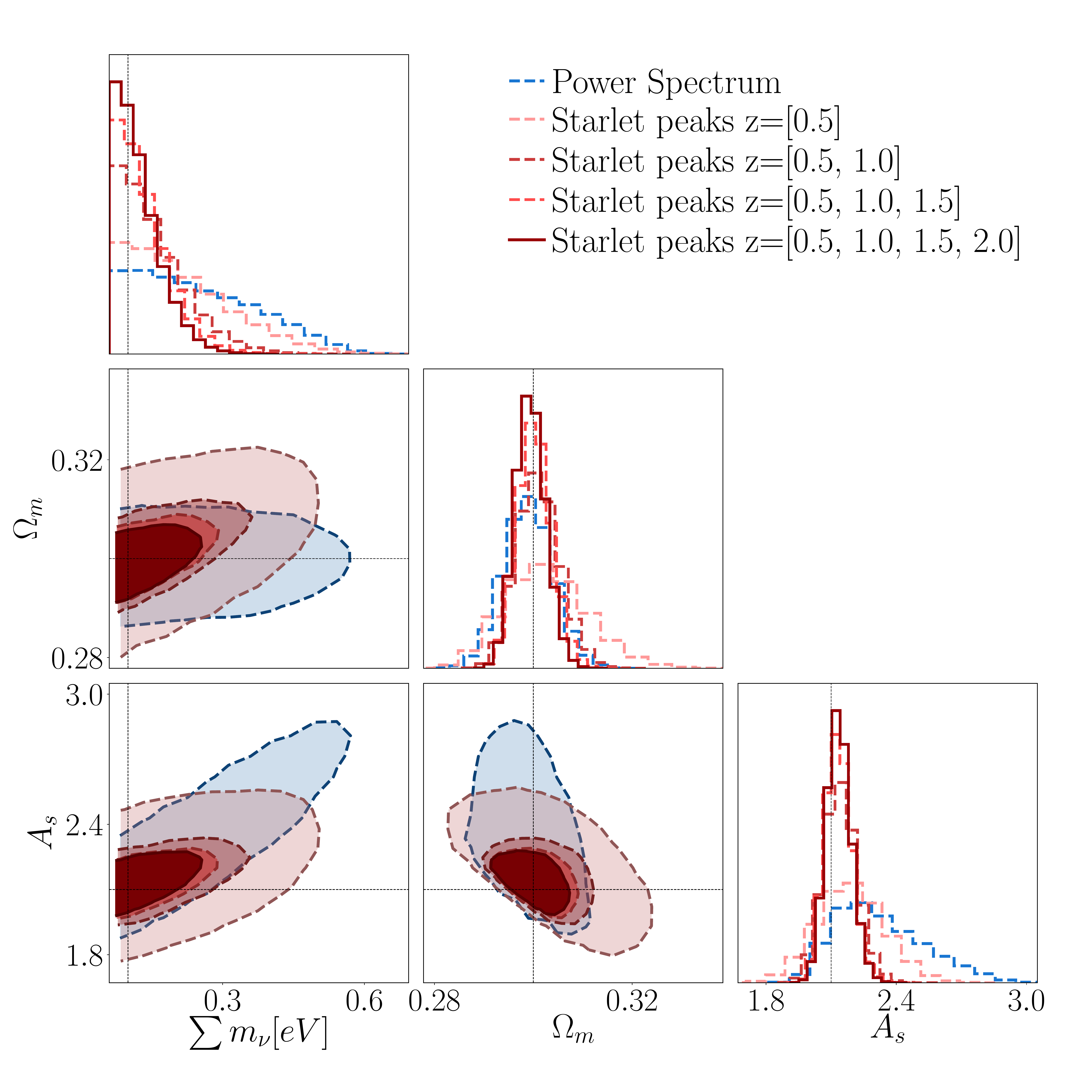}
  \caption{\label{fig:starlet_scales_contours}95 \% confidence contours using tomography with redshifts $z_s=[0.5,1.0,1.5,2.0]$ and corresponding galaxy number densities $n_\mathrm{gal}=[11.02, 11.90, 5.45, 1.45]$. The black dotted line is the fiducial model: $[\sum m_{\nu}, \Omega_{\rm m}, 10^9A_{\rm s}] = [0.1, 0.3, 2.1]$. \textbf{Left panel}: we show the impact of the different starlet scales and we prove there exists a minimum resolution $\theta_\mathrm{ker}=w_2=3.2$ arcmin that allows us to obtain constraints comparable to what is achieved with the full wavelet decomposition and that the information contained in the coarse map cannot be neglected. Starting from the first starlet scale alone $\theta_\mathrm{ker}=w_1=1.6$ arcmin (dashed big contours in light blue) we add scale by scale in light blue until $[w_1, w_2, w_3]$. In dark blue we show the constraints corresponding to $[w_2, w_3, c_4]$ that almost match with the constraints provided by the full starlet decomposition (in red). \textbf{Right panel}: we show here the constraints obtained by adding each tomographic redshift at the time: the dashed pink corresponds to the contours relative to $z_s=0.5$, the lighter red to $z_s=[0.5, 1.0]$, the darker pink to $z_s=[0.5, 1.0, 1.5]$ and the dark red to the full set of redshifts in the starlet case. We plot this against the power spectrum (blue contours) to show how each source redshift contribution to shrinks the contours and helps break the degeneracy with respect to the power spectrum.}\label{fig:starlet_scales_redshifts}
\end{figure*}

\medskip
\medskip

\appendix

\section{Physical interpretation}

Here we investigate how parameter constraint contours shrink by adding the tomographic redshift bins by one by one. In particular, in the right panel of Fig. \ref{fig:starlet_scales_redshifts} we show the information gain resulting from the addition of each tomographic redshift bin. In blue we show the power spectrum, and in darkening shades of red we plot contours for the starlet peaks as follows. Contours for source redshift $z_s=0.5$ are dashed pink, they are dashed darker pink for source redshift $z_s=[0.5, 1.0]$, and so on until reaching the dark red contours which are obtained by concatenating all source redshifts. As expected, the peaks contours show a different degeneracy direction from that of the power spectrum due to the higher-order information they contain. Indeed, the contours at $z_s=0.5$ already show different degeneracy compared to the power spectrum contours between $M_{\nu}$ and $\Omega_{\rm m}$ with a FoC=1.09, giving though larger marginalised constraints on $\Omega_{\rm m}$. Adding $z_s=1.0$ provides FoM that are more three times those for $z_s=0.5$ alone for the planes $(M_{\nu}, \Omega_{\rm m})$ and $(M_{\nu}, A_{\rm s})$ but increases the  correlation between $M_{\nu}$ and $\Omega_{\rm m}$ to FoC=1.17. Finally, the concatenation of the source redshifts $z_s=[0.5, 1.0, 1.5]$ further adds correlations between the two parameters (FoC=1.19) but shrinks the contours to almost reach those of the complete set of source redshifts.

\begin{acknowledgments}
\noindent VA wishes to thank Martin Kilbinger, Santiago Casas, Samuel Farrens, Nicolas Martinet, Niall Jeffrey, Jos\'e Manuel Zorrilla Matilla, Sebastian Rojas Gonzalez and Inneke Van Nieuwenhuyse for helpful discussions. VA acknowledges support by the Centre National d’Etudes Spatiales and the project Initiative d’Excellence (IdEx) of Université de Paris (ANR-18-IDEX-0001). We thank the Columbia Lensing group \citep{columbia_lensing} for making their simulations available. The creation of these simulations is supported through grants NSF AST-1210877, NSF AST-140041, and NASA ATP-80NSSC18K1093. We thank New Mexico State University (USA) and Instituto de Astrofisica de Andalucia CSIC (Spain) for hosting the Skies \& Universes site for cosmological simulation products.
\end{acknowledgments}

\appendix

\bibliographystyle{apsrev4-1}

\bibliography{biblio2}

\begin{thebibliography}{62}%
\makeatletter
\providecommand \@ifxundefined [1]{%
 \@ifx{#1\undefined}
}%
\providecommand \@ifnum [1]{%
 \ifnum #1\expandafter \@firstoftwo
 \else \expandafter \@secondoftwo
 \fi
}%
\providecommand \@ifx [1]{%
 \ifx #1\expandafter \@firstoftwo
 \else \expandafter \@secondoftwo
 \fi
}%
\providecommand \natexlab [1]{#1}%
\providecommand \enquote  [1]{``#1''}%
\providecommand \bibnamefont  [1]{#1}%
\providecommand \bibfnamefont [1]{#1}%
\providecommand \citenamefont [1]{#1}%
\providecommand \href@noop [0]{\@secondoftwo}%
\providecommand \href [0]{\begingroup \@sanitize@url \@href}%
\providecommand \@href[1]{\@@startlink{#1}\@@href}%
\providecommand \@@href[1]{\endgroup#1\@@endlink}%
\providecommand \@sanitize@url [0]{\catcode `\\12\catcode `\$12\catcode
  `\&12\catcode `\#12\catcode `\^12\catcode `\_12\catcode `\%12\relax}%
\providecommand \@@startlink[1]{}%
\providecommand \@@endlink[0]{}%
\providecommand \url  [0]{\begingroup\@sanitize@url \@url }%
\providecommand \@url [1]{\endgroup\@href {#1}{\urlprefix }}%
\providecommand \urlprefix  [0]{URL }%
\providecommand \Eprint [0]{\href }%
\providecommand \doibase [0]{http://dx.doi.org/}%
\providecommand \selectlanguage [0]{\@gobble}%
\providecommand \bibinfo  [0]{\@secondoftwo}%
\providecommand \bibfield  [0]{\@secondoftwo}%
\providecommand \translation [1]{[#1]}%
\providecommand \BibitemOpen [0]{}%
\providecommand \bibitemStop [0]{}%
\providecommand \bibitemNoStop [0]{.\EOS\space}%
\providecommand \EOS [0]{\spacefactor3000\relax}%
\providecommand \BibitemShut  [1]{\csname bibitem#1\endcsname}%
\let\auto@bib@innerbib\@empty
\bibitem [{\citenamefont {{Lesgourgues}}\ and\ \citenamefont
  {{Pastor}}(2006)}]{2006PhR...429..307L}%
  \BibitemOpen
  \bibfield  {author} {\bibinfo {author} {\bibfnamefont {J.}~\bibnamefont
  {{Lesgourgues}}}\ and\ \bibinfo {author} {\bibfnamefont {S.}~\bibnamefont
  {{Pastor}}},\ }\href {\doibase 10.1016/j.physrep.2006.04.001} {\bibfield
  {journal} {\bibinfo  {journal} {\physrep}\ }\textbf {\bibinfo {volume}
  {429}},\ \bibinfo {pages} {307} (\bibinfo {year} {2006})}\BibitemShut
  {NoStop}%
\bibitem [{\citenamefont {Peel}\ \emph {et~al.}(2019)\citenamefont {Peel},
  \citenamefont {Lalande}, \citenamefont {Starck}, \citenamefont {Pettorino},
  \citenamefont {Merten}, \citenamefont {Giocoli}, \citenamefont {Meneghetti},\
  and\ \citenamefont {Baldi}}]{PhysRevD.100.023508}%
  \BibitemOpen
  \bibfield  {author} {\bibinfo {author} {\bibfnamefont {A.}~\bibnamefont
  {Peel}}, \bibinfo {author} {\bibfnamefont {F.}~\bibnamefont {Lalande}},
  \bibinfo {author} {\bibfnamefont {J.-L.}\ \bibnamefont {Starck}}, \bibinfo
  {author} {\bibfnamefont {V.}~\bibnamefont {Pettorino}}, \bibinfo {author}
  {\bibfnamefont {J.}~\bibnamefont {Merten}}, \bibinfo {author} {\bibfnamefont
  {C.}~\bibnamefont {Giocoli}}, \bibinfo {author} {\bibfnamefont
  {M.}~\bibnamefont {Meneghetti}}, \ and\ \bibinfo {author} {\bibfnamefont
  {M.}~\bibnamefont {Baldi}},\ }\href {\doibase 10.1103/PhysRevD.100.023508}
  {\bibfield  {journal} {\bibinfo  {journal} {Phys. Rev. D}\ }\textbf {\bibinfo
  {volume} {100}},\ \bibinfo {pages} {023508} (\bibinfo {year}
  {2019})}\BibitemShut {NoStop}%
\bibitem [{\citenamefont {{Hagstotz, Steffen}}\ \emph
  {et~al.}(2019)\citenamefont {{Hagstotz, Steffen}}, \citenamefont {{Gronke,
  Max}}, \citenamefont {{Mota, David F.}},\ and\ \citenamefont {{Baldi,
  Marco}}}]{refId0}%
  \BibitemOpen
  \bibfield  {author} {\bibinfo {author} {\bibnamefont {{Hagstotz, Steffen}}},
  \bibinfo {author} {\bibnamefont {{Gronke, Max}}}, \bibinfo {author}
  {\bibnamefont {{Mota, David F.}}}, \ and\ \bibinfo {author} {\bibnamefont
  {{Baldi, Marco}}},\ }\href {\doibase 10.1051/0004-6361/201935213} {\bibfield
  {journal} {\bibinfo  {journal} {A\&A}\ }\textbf {\bibinfo {volume} {629}},\
  \bibinfo {pages} {A46} (\bibinfo {year} {2019})}\BibitemShut {NoStop}%
\bibitem [{\citenamefont {Merten}\ \emph {et~al.}(2019)\citenamefont {Merten},
  \citenamefont {Giocoli}, \citenamefont {Baldi}, \citenamefont {Meneghetti},
  \citenamefont {Peel}, \citenamefont {Lalande}, \citenamefont {Starck},\ and\
  \citenamefont {Pettorino}}]{10.1093/mnras/stz972}%
  \BibitemOpen
  \bibfield  {author} {\bibinfo {author} {\bibfnamefont {J.}~\bibnamefont
  {Merten}}, \bibinfo {author} {\bibfnamefont {C.}~\bibnamefont {Giocoli}},
  \bibinfo {author} {\bibfnamefont {M.}~\bibnamefont {Baldi}}, \bibinfo
  {author} {\bibfnamefont {M.}~\bibnamefont {Meneghetti}}, \bibinfo {author}
  {\bibfnamefont {A.}~\bibnamefont {Peel}}, \bibinfo {author} {\bibfnamefont
  {F.}~\bibnamefont {Lalande}}, \bibinfo {author} {\bibfnamefont {J.-L.}\
  \bibnamefont {Starck}}, \ and\ \bibinfo {author} {\bibfnamefont
  {V.}~\bibnamefont {Pettorino}},\ }\href {\doibase 10.1093/mnras/stz972}
  {\bibfield  {journal} {\bibinfo  {journal} {Monthly Notices of the Royal
  Astronomical Society}\ }\textbf {\bibinfo {volume} {487}},\ \bibinfo {pages}
  {104} (\bibinfo {year} {2019})}\BibitemShut {NoStop}%
\bibitem [{\citenamefont {Capozzi}\ \emph {et~al.}(2016)\citenamefont
  {Capozzi}, \citenamefont {Lisi}, \citenamefont {Marrone}, \citenamefont
  {Montanino},\ and\ \citenamefont {Palazzo}}]{CAPOZZI2016218}%
  \BibitemOpen
  \bibfield  {author} {\bibinfo {author} {\bibfnamefont {F.}~\bibnamefont
  {Capozzi}}, \bibinfo {author} {\bibfnamefont {E.}~\bibnamefont {Lisi}},
  \bibinfo {author} {\bibfnamefont {A.}~\bibnamefont {Marrone}}, \bibinfo
  {author} {\bibfnamefont {D.}~\bibnamefont {Montanino}}, \ and\ \bibinfo
  {author} {\bibfnamefont {A.}~\bibnamefont {Palazzo}},\ }\href {\doibase
  https://doi.org/10.1016/j.nuclphysb.2016.02.016} {\bibfield  {journal}
  {\bibinfo  {journal} {Nuclear Physics B}\ }\textbf {\bibinfo {volume}
  {908}},\ \bibinfo {pages} {218 } (\bibinfo {year} {2016})}\BibitemShut
  {NoStop}%
\bibitem [{\citenamefont {{Dealtry}}(2019)}]{2019arXiv190410206D}%
  \BibitemOpen
  \bibfield  {author} {\bibinfo {author} {\bibfnamefont {T.}~\bibnamefont
  {{Dealtry}}},\ }\href@noop {} {\bibfield  {journal} {\bibinfo  {journal}
  {arXiv e-prints}\ } (\bibinfo {year} {2019})},\ \Eprint
  {http://arxiv.org/abs/1904.10206} {arXiv:1904.10206} \BibitemShut {NoStop}%
\bibitem [{\citenamefont {{Planck Collaboration}}\ \emph
  {et~al.}(2020)\citenamefont {{Planck Collaboration}}, \citenamefont
  {{Aghanim, N.}}, \citenamefont {{Akrami, Y.}}, \citenamefont {{Ashdown, M.}},
  \citenamefont {{Aumont, J.}}, \citenamefont {{Baccigalupi, C.}},
  \citenamefont {{Ballardini, M.}}, \citenamefont {{Banday, A. J.}},
  \citenamefont {{Barreiro, R. B.}}, \citenamefont {{Bartolo, N.}},
  \citenamefont {{Basak, S.}}, \citenamefont {{Battye, R.}}, \citenamefont
  {{Benabed, K.}}, \citenamefont {{Bernard, J.-P.}} \emph
  {et~al.}}]{Planck2018}%
  \BibitemOpen
  \bibfield  {author} {\bibinfo {author} {\bibnamefont {{Planck
  Collaboration}}}, \bibinfo {author} {\bibnamefont {{Aghanim, N.}}}, \bibinfo
  {author} {\bibnamefont {{Akrami, Y.}}}, \bibinfo {author} {\bibnamefont
  {{Ashdown, M.}}}, \bibinfo {author} {\bibnamefont {{Aumont, J.}}}, \bibinfo
  {author} {\bibnamefont {{Baccigalupi, C.}}}, \bibinfo {author} {\bibnamefont
  {{Ballardini, M.}}}, \bibinfo {author} {\bibnamefont {{Banday, A. J.}}},
  \bibinfo {author} {\bibnamefont {{Barreiro, R. B.}}}, \bibinfo {author}
  {\bibnamefont {{Bartolo, N.}}}, \bibinfo {author} {\bibnamefont {{Basak,
  S.}}}, \bibinfo {author} {\bibnamefont {{Battye, R.}}}, \bibinfo {author}
  {\bibnamefont {{Benabed, K.}}}, \bibinfo {author} {\bibnamefont {{Bernard,
  J.-P.}}},  \emph {et~al.},\ }\href {\doibase 10.1051/0004-6361/201833910}
  {\bibfield  {journal} {\bibinfo  {journal} {A\&A}\ }\textbf {\bibinfo
  {volume} {641}},\ \bibinfo {pages} {A6} (\bibinfo {year} {2020})}\BibitemShut
  {NoStop}%
\bibitem [{\citenamefont {Heymans}\ \emph {et~al.}(2012)\citenamefont
  {Heymans}, \citenamefont {van Waerbeke}, \citenamefont {Miller},
  \citenamefont {Erben}, \citenamefont {Hildebrandt}, \citenamefont {Hoekstra},
  \citenamefont {Kitching}, \citenamefont {Mellier}, \citenamefont {Simon},
  \citenamefont {Bonnett}, \citenamefont {Coupon}, \citenamefont {Fu},
  \citenamefont {Harnois-Déraps}, \citenamefont {Hudson}, \citenamefont
  {Kilbinger}, \citenamefont {Kuijken}, \citenamefont {Rowe}, \citenamefont
  {Schrabback}, \citenamefont {Semboloni}, \citenamefont {van Uitert},
  \citenamefont {Vafaei},\ and\ \citenamefont
  {Velander}}]{10.1111/j.1365-2966.2012.21952.x}%
  \BibitemOpen
  \bibfield  {author} {\bibinfo {author} {\bibfnamefont {C.}~\bibnamefont
  {Heymans}}, \bibinfo {author} {\bibfnamefont {L.}~\bibnamefont {van
  Waerbeke}}, \bibinfo {author} {\bibfnamefont {L.}~\bibnamefont {Miller}},
  \bibinfo {author} {\bibfnamefont {T.}~\bibnamefont {Erben}}, \bibinfo
  {author} {\bibfnamefont {H.}~\bibnamefont {Hildebrandt}}, \bibinfo {author}
  {\bibfnamefont {H.}~\bibnamefont {Hoekstra}}, \bibinfo {author}
  {\bibfnamefont {T.~D.}\ \bibnamefont {Kitching}}, \bibinfo {author}
  {\bibfnamefont {Y.}~\bibnamefont {Mellier}}, \bibinfo {author} {\bibfnamefont
  {P.}~\bibnamefont {Simon}}, \bibinfo {author} {\bibfnamefont
  {C.}~\bibnamefont {Bonnett}}, \bibinfo {author} {\bibfnamefont
  {J.}~\bibnamefont {Coupon}}, \bibinfo {author} {\bibfnamefont
  {L.}~\bibnamefont {Fu}}, \bibinfo {author} {\bibfnamefont {J.}~\bibnamefont
  {Harnois-Déraps}}, \bibinfo {author} {\bibfnamefont {M.~J.}\ \bibnamefont
  {Hudson}}, \bibinfo {author} {\bibfnamefont {M.}~\bibnamefont {Kilbinger}},
  \bibinfo {author} {\bibfnamefont {K.}~\bibnamefont {Kuijken}}, \bibinfo
  {author} {\bibfnamefont {B.}~\bibnamefont {Rowe}}, \bibinfo {author}
  {\bibfnamefont {T.}~\bibnamefont {Schrabback}}, \bibinfo {author}
  {\bibfnamefont {E.}~\bibnamefont {Semboloni}}, \bibinfo {author}
  {\bibfnamefont {E.}~\bibnamefont {van Uitert}}, \bibinfo {author}
  {\bibfnamefont {S.}~\bibnamefont {Vafaei}}, \ and\ \bibinfo {author}
  {\bibfnamefont {M.}~\bibnamefont {Velander}},\ }\href {\doibase
  10.1111/j.1365-2966.2012.21952.x} {\bibfield  {journal} {\bibinfo  {journal}
  {Monthly Notices of the Royal Astronomical Society}\ }\textbf {\bibinfo
  {volume} {427}},\ \bibinfo {pages} {146} (\bibinfo {year}
  {2012})}\BibitemShut {NoStop}%
\bibitem [{\citenamefont {Heymans}(2020)}]{heymans2020kids1000}%
  \BibitemOpen
  \bibfield  {author} {\bibinfo {author} {\bibfnamefont {C.}~\bibnamefont
  {Heymans}},\ }\href@noop {} {} (\bibinfo {year} {2020}),\ \Eprint
  {http://arxiv.org/abs/2007.15632} {arXiv:2007.15632 [astro-ph.CO]}
  \BibitemShut {NoStop}%
\bibitem [{\citenamefont {Abbott}\ \emph {et~al.}(2018)\citenamefont {Abbott},
  \citenamefont {Abdalla}, \citenamefont {Alarcon}, \citenamefont
  {Aleksi\ifmmode~\acute{c}\else \'{c}\fi{}}, \citenamefont {Allam},
  \citenamefont {Allen}, \citenamefont {Amara}, \citenamefont {Annis},
  \citenamefont {Asorey}, \citenamefont {Avila} \emph
  {et~al.}}]{PhysRevD.98.043526}%
  \BibitemOpen
  \bibfield  {author} {\bibinfo {author} {\bibfnamefont {T.~M.~C.}\
  \bibnamefont {Abbott}}, \bibinfo {author} {\bibfnamefont {F.~B.}\
  \bibnamefont {Abdalla}}, \bibinfo {author} {\bibfnamefont {A.}~\bibnamefont
  {Alarcon}}, \bibinfo {author} {\bibfnamefont {J.}~\bibnamefont
  {Aleksi\ifmmode~\acute{c}\else \'{c}\fi{}}}, \bibinfo {author} {\bibfnamefont
  {S.}~\bibnamefont {Allam}}, \bibinfo {author} {\bibfnamefont
  {S.}~\bibnamefont {Allen}}, \bibinfo {author} {\bibfnamefont
  {A.}~\bibnamefont {Amara}}, \bibinfo {author} {\bibfnamefont
  {J.}~\bibnamefont {Annis}}, \bibinfo {author} {\bibfnamefont
  {J.}~\bibnamefont {Asorey}}, \bibinfo {author} {\bibnamefont {Avila}},  \emph
  {et~al.} (\bibinfo {collaboration} {Dark Energy Survey Collaboration 1}),\
  }\href {\doibase 10.1103/PhysRevD.98.043526} {\bibfield  {journal} {\bibinfo
  {journal} {Phys. Rev. D}\ }\textbf {\bibinfo {volume} {98}},\ \bibinfo
  {pages} {043526} (\bibinfo {year} {2018})}\BibitemShut {NoStop}%
\bibitem [{\citenamefont {Mandelbaum}\ \emph {et~al.}(2017)\citenamefont
  {Mandelbaum}, \citenamefont {Miyatake}, \citenamefont {Hamana} \emph
  {et~al.}}]{10.1093/pasj/psx130}%
  \BibitemOpen
  \bibfield  {author} {\bibinfo {author} {\bibfnamefont {R.}~\bibnamefont
  {Mandelbaum}}, \bibinfo {author} {\bibfnamefont {H.}~\bibnamefont
  {Miyatake}}, \bibinfo {author} {\bibfnamefont {T.}~\bibnamefont {Hamana}},
  \emph {et~al.},\ }\href {https://doi.org/10.1093/pasj/psx130} {\bibfield
  {journal} {\bibinfo  {journal} {Publications of the Astronomical Society of
  Japan}\ }\textbf {\bibinfo {volume} {70}} (\bibinfo {year} {2017})},\
  \bibinfo {note} {s25}\BibitemShut {NoStop}%
\bibitem [{\citenamefont {Hikage}\ \emph {et~al.}(2019)\citenamefont {Hikage},
  \citenamefont {Oguri}, \citenamefont {Hamana} \emph
  {et~al.}}]{10.1093/pasj/psz010}%
  \BibitemOpen
  \bibfield  {author} {\bibinfo {author} {\bibfnamefont {C.}~\bibnamefont
  {Hikage}}, \bibinfo {author} {\bibfnamefont {M.}~\bibnamefont {Oguri}},
  \bibinfo {author} {\bibfnamefont {T.}~\bibnamefont {Hamana}},  \emph
  {et~al.},\ }\href {https://doi.org/10.1093/pasj/psz010} {\bibfield  {journal}
  {\bibinfo  {journal} {Publications of the Astronomical Society of Japan}\
  }\textbf {\bibinfo {volume} {71}} (\bibinfo {year} {2019})},\ \bibinfo {note}
  {43}\BibitemShut {NoStop}%
\bibitem [{\citenamefont {Laureijs}\ \emph {et~al.}(2011)\citenamefont
  {Laureijs}, \citenamefont {Amiaux}, \citenamefont {Arduini}, \citenamefont
  {Auguères}, \citenamefont {Brinchmann}, \citenamefont {Cole}, \citenamefont
  {Cropper}, \citenamefont {Dabin}, \citenamefont {Duvet}, \citenamefont
  {Ealet}, \citenamefont {Garilli}, \citenamefont {Gondoin}, \citenamefont
  {Guzzo}, \citenamefont {Hoar}, \citenamefont {Hoekstra}, \citenamefont
  {Holmes}, \citenamefont {Kitching} \emph {et~al.}}]{laureijs2011euclid}%
  \BibitemOpen
  \bibfield  {author} {\bibinfo {author} {\bibfnamefont {R.}~\bibnamefont
  {Laureijs}}, \bibinfo {author} {\bibfnamefont {J.}~\bibnamefont {Amiaux}},
  \bibinfo {author} {\bibfnamefont {S.}~\bibnamefont {Arduini}}, \bibinfo
  {author} {\bibfnamefont {J.~L.}\ \bibnamefont {Auguères}}, \bibinfo {author}
  {\bibfnamefont {J.}~\bibnamefont {Brinchmann}}, \bibinfo {author}
  {\bibfnamefont {R.}~\bibnamefont {Cole}}, \bibinfo {author} {\bibfnamefont
  {M.}~\bibnamefont {Cropper}}, \bibinfo {author} {\bibfnamefont
  {C.}~\bibnamefont {Dabin}}, \bibinfo {author} {\bibfnamefont
  {L.}~\bibnamefont {Duvet}}, \bibinfo {author} {\bibfnamefont
  {A.}~\bibnamefont {Ealet}}, \bibinfo {author} {\bibfnamefont
  {B.}~\bibnamefont {Garilli}}, \bibinfo {author} {\bibfnamefont
  {P.}~\bibnamefont {Gondoin}}, \bibinfo {author} {\bibfnamefont
  {L.}~\bibnamefont {Guzzo}}, \bibinfo {author} {\bibfnamefont
  {J.}~\bibnamefont {Hoar}}, \bibinfo {author} {\bibfnamefont {H.}~\bibnamefont
  {Hoekstra}}, \bibinfo {author} {\bibfnamefont {R.}~\bibnamefont {Holmes}},
  \bibinfo {author} {\bibfnamefont {T.}~\bibnamefont {Kitching}},  \emph
  {et~al.},\ }\href@noop {} {} (\bibinfo {year} {2011}),\ \Eprint
  {http://arxiv.org/abs/1110.3193} {arXiv:1110.3193 [astro-ph.CO]} \BibitemShut
  {NoStop}%
\bibitem [{\citenamefont {Rizzato}\ \emph {et~al.}(2019)\citenamefont
  {Rizzato}, \citenamefont {Benabed}, \citenamefont {Bernardeau},\ and\
  \citenamefont {Lacasa}}]{10.1093/mnras/stz2862}%
  \BibitemOpen
  \bibfield  {author} {\bibinfo {author} {\bibfnamefont {M.}~\bibnamefont
  {Rizzato}}, \bibinfo {author} {\bibfnamefont {K.}~\bibnamefont {Benabed}},
  \bibinfo {author} {\bibfnamefont {F.}~\bibnamefont {Bernardeau}}, \ and\
  \bibinfo {author} {\bibfnamefont {F.}~\bibnamefont {Lacasa}},\ }\href
  {\doibase 10.1093/mnras/stz2862} {\bibfield  {journal} {\bibinfo  {journal}
  {Monthly Notices of the Royal Astronomical Society}\ }\textbf {\bibinfo
  {volume} {490}},\ \bibinfo {pages} {4688} (\bibinfo {year}
  {2019})}\BibitemShut {NoStop}%
\bibitem [{\citenamefont {Kayo}\ and\ \citenamefont
  {Takada}(2013)}]{kayo2013cosmological}%
  \BibitemOpen
  \bibfield  {author} {\bibinfo {author} {\bibfnamefont {I.}~\bibnamefont
  {Kayo}}\ and\ \bibinfo {author} {\bibfnamefont {M.}~\bibnamefont {Takada}},\
  }\href@noop {} {} (\bibinfo {year} {2013}),\ \Eprint
  {http://arxiv.org/abs/1306.4684} {arXiv:1306.4684} \BibitemShut {NoStop}%
\bibitem [{\citenamefont {Petri}\ \emph {et~al.}(2013)\citenamefont {Petri},
  \citenamefont {Haiman}, \citenamefont {Hui}, \citenamefont {May},\ and\
  \citenamefont {Kratochvil}}]{PhysRevD.88.123002}%
  \BibitemOpen
  \bibfield  {author} {\bibinfo {author} {\bibfnamefont {A.}~\bibnamefont
  {Petri}}, \bibinfo {author} {\bibfnamefont {Z.}~\bibnamefont {Haiman}},
  \bibinfo {author} {\bibfnamefont {L.}~\bibnamefont {Hui}}, \bibinfo {author}
  {\bibfnamefont {M.}~\bibnamefont {May}}, \ and\ \bibinfo {author}
  {\bibfnamefont {J.~M.}\ \bibnamefont {Kratochvil}},\ }\href {\doibase
  10.1103/PhysRevD.88.123002} {\bibfield  {journal} {\bibinfo  {journal} {Phys.
  Rev. D}\ }\textbf {\bibinfo {volume} {88}},\ \bibinfo {pages} {123002}
  (\bibinfo {year} {2013})}\BibitemShut {NoStop}%
\bibitem [{\citenamefont {Kratochvil}\ \emph {et~al.}(2012)\citenamefont
  {Kratochvil}, \citenamefont {Lim}, \citenamefont {Wang}, \citenamefont
  {Haiman}, \citenamefont {May},\ and\ \citenamefont
  {Huffenberger}}]{PhysRevD.85.103513}%
  \BibitemOpen
  \bibfield  {author} {\bibinfo {author} {\bibfnamefont {J.~M.}\ \bibnamefont
  {Kratochvil}}, \bibinfo {author} {\bibfnamefont {E.~A.}\ \bibnamefont {Lim}},
  \bibinfo {author} {\bibfnamefont {S.}~\bibnamefont {Wang}}, \bibinfo {author}
  {\bibfnamefont {Z.}~\bibnamefont {Haiman}}, \bibinfo {author} {\bibfnamefont
  {M.}~\bibnamefont {May}}, \ and\ \bibinfo {author} {\bibfnamefont
  {K.}~\bibnamefont {Huffenberger}},\ }\href {\doibase
  10.1103/PhysRevD.85.103513} {\bibfield  {journal} {\bibinfo  {journal} {Phys.
  Rev. D}\ }\textbf {\bibinfo {volume} {85}},\ \bibinfo {pages} {103513}
  (\bibinfo {year} {2012})}\BibitemShut {NoStop}%
\bibitem [{\citenamefont {Kacprzak}\ \emph {et~al.}(2016)\citenamefont
  {Kacprzak}, \citenamefont {Kirk}, \citenamefont {Friedrich}, \citenamefont
  {Amara}, \citenamefont {Refregier} \emph {et~al.}}]{Kacprzak2016}%
  \BibitemOpen
  \bibfield  {author} {\bibinfo {author} {\bibfnamefont {T.}~\bibnamefont
  {Kacprzak}}, \bibinfo {author} {\bibfnamefont {D.}~\bibnamefont {Kirk}},
  \bibinfo {author} {\bibfnamefont {O.}~\bibnamefont {Friedrich}}, \bibinfo
  {author} {\bibfnamefont {A.}~\bibnamefont {Amara}}, \bibinfo {author}
  {\bibfnamefont {A.}~\bibnamefont {Refregier}},  \emph {et~al.},\ }\href
  {\doibase 10.1093/mnras/stw2070} {\bibfield  {journal} {\bibinfo  {journal}
  {Monthly Notices of the Royal Astronomical Society}\ }\textbf {\bibinfo
  {volume} {463}},\ \bibinfo {pages} {3653} (\bibinfo {year}
  {2016})}\BibitemShut {NoStop}%
\bibitem [{\citenamefont {{Lin, Chieh-An}}\ and\ \citenamefont {{Kilbinger,
  Martin}}(2015)}]{Linc2015kilb}%
  \BibitemOpen
  \bibfield  {author} {\bibinfo {author} {\bibnamefont {{Lin, Chieh-An}}}\ and\
  \bibinfo {author} {\bibnamefont {{Kilbinger, Martin}}},\ }\href {\doibase
  10.1051/0004-6361/201526659} {\bibfield  {journal} {\bibinfo  {journal}
  {A\&A}\ }\textbf {\bibinfo {volume} {583}},\ \bibinfo {pages} {A70} (\bibinfo
  {year} {2015})}\BibitemShut {NoStop}%
\bibitem [{\citenamefont {{Peel, Austin}}\ \emph {et~al.}(2017)\citenamefont
  {{Peel, Austin}}, \citenamefont {{Lin, Chieh-An}}, \citenamefont {{Lanusse,
  Fran\c{c}ois}}, \citenamefont {{Leonard, Adrienne}}, \citenamefont {{Starck,
  Jean-Luc}},\ and\ \citenamefont {{Kilbinger, Martin}}}]{Peel2017Linc}%
  \BibitemOpen
  \bibfield  {author} {\bibinfo {author} {\bibnamefont {{Peel, Austin}}},
  \bibinfo {author} {\bibnamefont {{Lin, Chieh-An}}}, \bibinfo {author}
  {\bibnamefont {{Lanusse, Fran\c{c}ois}}}, \bibinfo {author} {\bibnamefont
  {{Leonard, Adrienne}}}, \bibinfo {author} {\bibnamefont {{Starck,
  Jean-Luc}}}, \ and\ \bibinfo {author} {\bibnamefont {{Kilbinger, Martin}}},\
  }\href {\doibase 10.1051/0004-6361/201629928} {\bibfield  {journal} {\bibinfo
   {journal} {A\&A}\ }\textbf {\bibinfo {volume} {599}},\ \bibinfo {pages}
  {A79} (\bibinfo {year} {2017})}\BibitemShut {NoStop}%
\bibitem [{\citenamefont {{Martinet, Nicolas}}\ \emph
  {et~al.}(2015)\citenamefont {{Martinet, Nicolas}}, \citenamefont {{Bartlett,
  James G.}}, \citenamefont {{Kiessling, Alina}},\ and\ \citenamefont
  {{Sartoris, Barbara}}}]{Martinet2015}%
  \BibitemOpen
  \bibfield  {author} {\bibinfo {author} {\bibnamefont {{Martinet, Nicolas}}},
  \bibinfo {author} {\bibnamefont {{Bartlett, James G.}}}, \bibinfo {author}
  {\bibnamefont {{Kiessling, Alina}}}, \ and\ \bibinfo {author} {\bibnamefont
  {{Sartoris, Barbara}}},\ }\href {\doibase 10.1051/0004-6361/201425164}
  {\bibfield  {journal} {\bibinfo  {journal} {A\&A}\ }\textbf {\bibinfo
  {volume} {581}},\ \bibinfo {pages} {A101} (\bibinfo {year}
  {2015})}\BibitemShut {NoStop}%
\bibitem [{\citenamefont {Shan}\ \emph {et~al.}(2017)\citenamefont {Shan} \emph
  {et~al.}}]{Shan2017}%
  \BibitemOpen
  \bibfield  {author} {\bibinfo {author} {\bibfnamefont {H.}~\bibnamefont
  {Shan}} \emph {et~al.},\ }\href {\doibase 10.1093/mnras/stx2837} {\bibfield
  {journal} {\bibinfo  {journal} {Monthly Notices of the Royal Astronomical
  Society}\ }\textbf {\bibinfo {volume} {474}},\ \bibinfo {pages} {1116}
  (\bibinfo {year} {2017})}\BibitemShut {NoStop}%
\bibitem [{\citenamefont {Martinet}\ \emph {et~al.}(2017)\citenamefont
  {Martinet}, \citenamefont {Schneider}, \citenamefont {Hildebrandt},
  \citenamefont {Shan}, \citenamefont {Asgari}, \citenamefont {Dietrich} \emph
  {et~al.}}]{MartinetSchneider}%
  \BibitemOpen
  \bibfield  {author} {\bibinfo {author} {\bibfnamefont {N.}~\bibnamefont
  {Martinet}}, \bibinfo {author} {\bibfnamefont {P.}~\bibnamefont {Schneider}},
  \bibinfo {author} {\bibfnamefont {H.}~\bibnamefont {Hildebrandt}}, \bibinfo
  {author} {\bibfnamefont {H.}~\bibnamefont {Shan}}, \bibinfo {author}
  {\bibfnamefont {M.}~\bibnamefont {Asgari}}, \bibinfo {author} {\bibfnamefont
  {J.~P.}\ \bibnamefont {Dietrich}},  \emph {et~al.},\ }\href {\doibase
  10.1093/mnras/stx2793} {\bibfield  {journal} {\bibinfo  {journal} {Monthly
  Notices of the Royal Astronomical Society}\ }\textbf {\bibinfo {volume}
  {474}},\ \bibinfo {pages} {712} (\bibinfo {year} {2017})}\BibitemShut
  {NoStop}%
\bibitem [{\citenamefont {Fluri}\ \emph {et~al.}(2018)\citenamefont {Fluri},
  \citenamefont {Kacprzak}, \citenamefont {Sgier}, \citenamefont {Refregier},\
  and\ \citenamefont {Amara}}]{Fluri_2018}%
  \BibitemOpen
  \bibfield  {author} {\bibinfo {author} {\bibfnamefont {J.}~\bibnamefont
  {Fluri}}, \bibinfo {author} {\bibfnamefont {T.}~\bibnamefont {Kacprzak}},
  \bibinfo {author} {\bibfnamefont {R.}~\bibnamefont {Sgier}}, \bibinfo
  {author} {\bibfnamefont {A.}~\bibnamefont {Refregier}}, \ and\ \bibinfo
  {author} {\bibfnamefont {A.}~\bibnamefont {Amara}},\ }\href {\doibase
  10.1088/1475-7516/2018/10/051} {\bibfield  {journal} {\bibinfo  {journal}
  {Journal of Cosmology and Astroparticle Physics}\ }\textbf {\bibinfo {volume}
  {2018}},\ \bibinfo {pages} {051} (\bibinfo {year} {2018})}\BibitemShut
  {NoStop}%
\bibitem [{\citenamefont {Zürcher}\ \emph {et~al.}(2020)\citenamefont
  {Zürcher}, \citenamefont {Fluri}, \citenamefont {Sgier}, \citenamefont
  {Kacprzak},\ and\ \citenamefont {Refregier}}]{zurcher2020cosmological}%
  \BibitemOpen
  \bibfield  {author} {\bibinfo {author} {\bibfnamefont {D.}~\bibnamefont
  {Zürcher}}, \bibinfo {author} {\bibfnamefont {J.}~\bibnamefont {Fluri}},
  \bibinfo {author} {\bibfnamefont {R.}~\bibnamefont {Sgier}}, \bibinfo
  {author} {\bibfnamefont {T.}~\bibnamefont {Kacprzak}}, \ and\ \bibinfo
  {author} {\bibfnamefont {A.}~\bibnamefont {Refregier}},\ }\href@noop {} {}
  (\bibinfo {year} {2020}),\ \Eprint {http://arxiv.org/abs/2006.12506}
  {arXiv:2006.12506 [astro-ph.CO]} \BibitemShut {NoStop}%
\bibitem [{\citenamefont {Petri}(2016)}]{PETRI201673}%
  \BibitemOpen
  \bibfield  {author} {\bibinfo {author} {\bibfnamefont {A.}~\bibnamefont
  {Petri}},\ }\href {\doibase https://doi.org/10.1016/j.ascom.2016.06.001}
  {\bibfield  {journal} {\bibinfo  {journal} {Astronomy and Computing}\
  }\textbf {\bibinfo {volume} {17}},\ \bibinfo {pages} {73 } (\bibinfo {year}
  {2016})}\BibitemShut {NoStop}%
\bibitem [{\citenamefont {Liu}\ \emph {et~al.}(2018)\citenamefont {Liu},
  \citenamefont {Bird}, \citenamefont {Matilla}, \citenamefont {Hill},
  \citenamefont {Haiman}, \citenamefont {Madhavacheril}, \citenamefont
  {Petri},\ and\ \citenamefont {Spergel}}]{Liu_2018}%
  \BibitemOpen
  \bibfield  {author} {\bibinfo {author} {\bibfnamefont {J.}~\bibnamefont
  {Liu}}, \bibinfo {author} {\bibfnamefont {S.}~\bibnamefont {Bird}}, \bibinfo
  {author} {\bibfnamefont {J.~M.~Z.}\ \bibnamefont {Matilla}}, \bibinfo
  {author} {\bibfnamefont {J.~C.}\ \bibnamefont {Hill}}, \bibinfo {author}
  {\bibfnamefont {Z.}~\bibnamefont {Haiman}}, \bibinfo {author} {\bibfnamefont
  {M.~S.}\ \bibnamefont {Madhavacheril}}, \bibinfo {author} {\bibfnamefont
  {A.}~\bibnamefont {Petri}}, \ and\ \bibinfo {author} {\bibfnamefont {D.~N.}\
  \bibnamefont {Spergel}},\ }\href {\doibase 10.1088/1475-7516/2018/03/049}
  {\bibfield  {journal} {\bibinfo  {journal} {Journal of Cosmology and
  Astroparticle Physics}\ }\textbf {\bibinfo {volume} {2018}},\ \bibinfo
  {pages} {049} (\bibinfo {year} {2018})}\BibitemShut {NoStop}%
\bibitem [{\citenamefont {Liu}\ and\ \citenamefont
  {Madhavacheril}(2019)}]{PhysRevD.99.083508}%
  \BibitemOpen
  \bibfield  {author} {\bibinfo {author} {\bibfnamefont {J.}~\bibnamefont
  {Liu}}\ and\ \bibinfo {author} {\bibfnamefont {M.~S.}\ \bibnamefont
  {Madhavacheril}},\ }\href {\doibase 10.1103/PhysRevD.99.083508} {\bibfield
  {journal} {\bibinfo  {journal} {Phys. Rev. D}\ }\textbf {\bibinfo {volume}
  {99}},\ \bibinfo {pages} {083508} (\bibinfo {year} {2019})}\BibitemShut
  {NoStop}%
\bibitem [{\citenamefont {Marques}\ \emph {et~al.}(2019)\citenamefont
  {Marques}, \citenamefont {Liu}, \citenamefont {Matilla}, \citenamefont
  {Haiman}, \citenamefont {Bernui},\ and\ \citenamefont
  {Novaes}}]{Marques_2019}%
  \BibitemOpen
  \bibfield  {author} {\bibinfo {author} {\bibfnamefont {G.~A.}\ \bibnamefont
  {Marques}}, \bibinfo {author} {\bibfnamefont {J.}~\bibnamefont {Liu}},
  \bibinfo {author} {\bibfnamefont {J.~M.~Z.}\ \bibnamefont {Matilla}},
  \bibinfo {author} {\bibfnamefont {Z.}~\bibnamefont {Haiman}}, \bibinfo
  {author} {\bibfnamefont {A.}~\bibnamefont {Bernui}}, \ and\ \bibinfo {author}
  {\bibfnamefont {C.~P.}\ \bibnamefont {Novaes}},\ }\href {\doibase
  10.1088/1475-7516/2019/06/019} {\bibfield  {journal} {\bibinfo  {journal}
  {Journal of Cosmology and Astroparticle Physics}\ }\textbf {\bibinfo {volume}
  {2019}},\ \bibinfo {pages} {019} (\bibinfo {year} {2019})}\BibitemShut
  {NoStop}%
\bibitem [{\citenamefont {Coulton}\ \emph {et~al.}(2019)\citenamefont
  {Coulton}, \citenamefont {Liu}, \citenamefont {Madhavacheril}, \citenamefont
  {Böhm},\ and\ \citenamefont {Spergel}}]{Coulton_2019}%
  \BibitemOpen
  \bibfield  {author} {\bibinfo {author} {\bibfnamefont {W.~R.}\ \bibnamefont
  {Coulton}}, \bibinfo {author} {\bibfnamefont {J.}~\bibnamefont {Liu}},
  \bibinfo {author} {\bibfnamefont {M.~S.}\ \bibnamefont {Madhavacheril}},
  \bibinfo {author} {\bibfnamefont {V.}~\bibnamefont {Böhm}}, \ and\ \bibinfo
  {author} {\bibfnamefont {D.~N.}\ \bibnamefont {Spergel}},\ }\href {\doibase
  10.1088/1475-7516/2019/05/043} {\bibfield  {journal} {\bibinfo  {journal}
  {Journal of Cosmology and Astroparticle Physics}\ }\textbf {\bibinfo {volume}
  {2019}},\ \bibinfo {pages} {043} (\bibinfo {year} {2019})}\BibitemShut
  {NoStop}%
\bibitem [{\citenamefont {Collaboration}\ \emph {et~al.}(2009)\citenamefont
  {Collaboration}, \citenamefont {Abell}, \citenamefont {Allison},
  \citenamefont {Anderson}, \citenamefont {Andrew}, \citenamefont {Angel},
  \citenamefont {Armus}, \citenamefont {Arnett}, \citenamefont {Asztalos},
  \citenamefont {Axelrod}, \citenamefont {Bailey}, \citenamefont {Ballantyne},
  \citenamefont {Bankert}, \citenamefont {Barkhouse}, \citenamefont {Barr},
  \citenamefont {Barrientos}, \citenamefont {Barth} \emph
  {et~al.}}]{lsstsciencecollaboration2009lsst}%
  \BibitemOpen
  \bibfield  {author} {\bibinfo {author} {\bibfnamefont {L.~S.}\ \bibnamefont
  {Collaboration}}, \bibinfo {author} {\bibfnamefont {P.~A.}\ \bibnamefont
  {Abell}}, \bibinfo {author} {\bibfnamefont {J.}~\bibnamefont {Allison}},
  \bibinfo {author} {\bibfnamefont {S.~F.}\ \bibnamefont {Anderson}}, \bibinfo
  {author} {\bibfnamefont {J.~R.}\ \bibnamefont {Andrew}}, \bibinfo {author}
  {\bibfnamefont {J.~R.~P.}\ \bibnamefont {Angel}}, \bibinfo {author}
  {\bibfnamefont {L.}~\bibnamefont {Armus}}, \bibinfo {author} {\bibfnamefont
  {D.}~\bibnamefont {Arnett}}, \bibinfo {author} {\bibfnamefont {S.~J.}\
  \bibnamefont {Asztalos}}, \bibinfo {author} {\bibfnamefont {T.~S.}\
  \bibnamefont {Axelrod}}, \bibinfo {author} {\bibfnamefont {S.}~\bibnamefont
  {Bailey}}, \bibinfo {author} {\bibfnamefont {D.~R.}\ \bibnamefont
  {Ballantyne}}, \bibinfo {author} {\bibfnamefont {J.~R.}\ \bibnamefont
  {Bankert}}, \bibinfo {author} {\bibfnamefont {W.~A.}\ \bibnamefont
  {Barkhouse}}, \bibinfo {author} {\bibfnamefont {J.~D.}\ \bibnamefont {Barr}},
  \bibinfo {author} {\bibfnamefont {L.~F.}\ \bibnamefont {Barrientos}},
  \bibinfo {author} {\bibfnamefont {A.~J.}\ \bibnamefont {Barth}},  \emph
  {et~al.},\ }\href@noop {} {} (\bibinfo {year} {2009}),\ \Eprint
  {http://arxiv.org/abs/0912.0201} {arXiv:0912.0201} \BibitemShut {NoStop}%
\bibitem [{\citenamefont {Li}\ \emph {et~al.}(2019)\citenamefont {Li},
  \citenamefont {Liu}, \citenamefont {Matilla},\ and\ \citenamefont
  {Coulton}}]{PhysRevD.99.063527}%
  \BibitemOpen
  \bibfield  {author} {\bibinfo {author} {\bibfnamefont {Z.}~\bibnamefont
  {Li}}, \bibinfo {author} {\bibfnamefont {J.}~\bibnamefont {Liu}}, \bibinfo
  {author} {\bibfnamefont {J.~M.~Z.}\ \bibnamefont {Matilla}}, \ and\ \bibinfo
  {author} {\bibfnamefont {W.~R.}\ \bibnamefont {Coulton}},\ }\href {\doibase
  10.1103/PhysRevD.99.063527} {\bibfield  {journal} {\bibinfo  {journal} {Phys.
  Rev. D}\ }\textbf {\bibinfo {volume} {99}},\ \bibinfo {pages} {063527}
  (\bibinfo {year} {2019})}\BibitemShut {NoStop}%
\bibitem [{\citenamefont {Starck}\ \emph {et~al.}(2010)\citenamefont {Starck},
  \citenamefont {Murtagh},\ and\ \citenamefont
  {Fadili}}]{starck_murtagh_fadili_2010}%
  \BibitemOpen
  \bibfield  {author} {\bibinfo {author} {\bibfnamefont {J.-L.}\ \bibnamefont
  {Starck}}, \bibinfo {author} {\bibfnamefont {F.}~\bibnamefont {Murtagh}}, \
  and\ \bibinfo {author} {\bibfnamefont {J.~M.}\ \bibnamefont {Fadili}},\
  }\href {\doibase 10.1017/CBO9780511730344} {\emph {\bibinfo {title} {Sparse
  Image and Signal Processing: Wavelets, Curvelets, Morphological Diversity}}}\
  (\bibinfo  {publisher} {Cambridge University Press},\ \bibinfo {year}
  {2010})\BibitemShut {NoStop}%
\bibitem [{\citenamefont {{Lin, Chieh-An}}\ \emph {et~al.}(2016)\citenamefont
  {{Lin, Chieh-An}}, \citenamefont {{Kilbinger, Martin}},\ and\ \citenamefont
  {{Pires, Sandrine}}}]{Linc2016}%
  \BibitemOpen
  \bibfield  {author} {\bibinfo {author} {\bibnamefont {{Lin, Chieh-An}}},
  \bibinfo {author} {\bibnamefont {{Kilbinger, Martin}}}, \ and\ \bibinfo
  {author} {\bibnamefont {{Pires, Sandrine}}},\ }\href {\doibase
  10.1051/0004-6361/201628565} {\bibfield  {journal} {\bibinfo  {journal}
  {A\&A}\ }\textbf {\bibinfo {volume} {593}},\ \bibinfo {pages} {A88} (\bibinfo
  {year} {2016})}\BibitemShut {NoStop}%
\bibitem [{\citenamefont {Kilbinger}(2015)}]{Kilbinger_2015}%
  \BibitemOpen
  \bibfield  {author} {\bibinfo {author} {\bibfnamefont {M.}~\bibnamefont
  {Kilbinger}},\ }\href {\doibase 10.1088/0034-4885/78/8/086901} {\bibfield
  {journal} {\bibinfo  {journal} {Reports on Progress in Physics}\ }\textbf
  {\bibinfo {volume} {78}},\ \bibinfo {pages} {086901} (\bibinfo {year}
  {2015})}\BibitemShut {NoStop}%
\bibitem [{\citenamefont {{Schneider}}\ \emph {et~al.}(1992)\citenamefont
  {{Schneider}}, \citenamefont {{Ehlers}},\ and\ \citenamefont
  {{Falco}}}]{1992grle.book.....S}%
  \BibitemOpen
  \bibfield  {author} {\bibinfo {author} {\bibfnamefont {P.}~\bibnamefont
  {{Schneider}}}, \bibinfo {author} {\bibfnamefont {J.}~\bibnamefont
  {{Ehlers}}}, \ and\ \bibinfo {author} {\bibfnamefont {E.~E.}\ \bibnamefont
  {{Falco}}},\ }\href {\doibase 10.1007/978-3-662-03758-4} {\emph {\bibinfo
  {title} {{Gravitational Lenses}}}}\ (\bibinfo {year} {1992})\BibitemShut
  {NoStop}%
\bibitem [{col()}]{columbia_lensing}%
  \BibitemOpen
  \href@noop {} {}\bibinfo {note}
  {\url{http://columbialensing.org/}}\BibitemShut {NoStop}%
\bibitem [{\citenamefont {{Kaiser}}\ and\ \citenamefont
  {{Squires}}(1993)}]{1993ApJ...404..441K}%
  \BibitemOpen
  \bibfield  {author} {\bibinfo {author} {\bibfnamefont {N.}~\bibnamefont
  {{Kaiser}}}\ and\ \bibinfo {author} {\bibfnamefont {G.}~\bibnamefont
  {{Squires}}},\ }\href {\doibase 10.1086/172297} {\bibfield  {journal}
  {\bibinfo  {journal} {\apj}\ }\textbf {\bibinfo {volume} {404}},\ \bibinfo
  {pages} {441} (\bibinfo {year} {1993})}\BibitemShut {NoStop}%
\bibitem [{\citenamefont {{Euclid Collaboration}}\ \emph
  {et~al.}(2020)\citenamefont {{Euclid Collaboration}}, \citenamefont
  {{Blanchard, A.}}, \citenamefont {{Camera, S.}}, \citenamefont {{Carbone,
  C.}}, \citenamefont {{Cardone, V. F.}}, \citenamefont {{Casas, S.}},
  \citenamefont {{Clesse, S.}}, \citenamefont {{Ili\'{}c, S.}}, \citenamefont
  {{Kilbinger, M.}}, \citenamefont {{Kitching, T.}}, \citenamefont {{Kunz,
  M.}}, \citenamefont {{Lacasa, F.}}, \citenamefont {{Linder, E.}},
  \citenamefont {{Majerotto, E.}}, \citenamefont {{Markovic, K.}},
  \citenamefont {{Martinelli, M.}}, \citenamefont {{Pettorino, V.}},
  \citenamefont {{Pourtsidou, A.}}, \citenamefont {{Sakr, Z.}}, \citenamefont
  {{S\'anchez, A. G.}}, \citenamefont {{Sapone, D.}}, \citenamefont {{Tutusaus,
  I.}} \emph {et~al.}}]{EuclidForecast2020}%
  \BibitemOpen
  \bibfield  {author} {\bibinfo {author} {\bibnamefont {{Euclid
  Collaboration}}}, \bibinfo {author} {\bibnamefont {{Blanchard, A.}}},
  \bibinfo {author} {\bibnamefont {{Camera, S.}}}, \bibinfo {author}
  {\bibnamefont {{Carbone, C.}}}, \bibinfo {author} {\bibnamefont {{Cardone, V.
  F.}}}, \bibinfo {author} {\bibnamefont {{Casas, S.}}}, \bibinfo {author}
  {\bibnamefont {{Clesse, S.}}}, \bibinfo {author} {\bibnamefont {{Ili\'{}c,
  S.}}}, \bibinfo {author} {\bibnamefont {{Kilbinger, M.}}}, \bibinfo {author}
  {\bibnamefont {{Kitching, T.}}}, \bibinfo {author} {\bibnamefont {{Kunz,
  M.}}}, \bibinfo {author} {\bibnamefont {{Lacasa, F.}}}, \bibinfo {author}
  {\bibnamefont {{Linder, E.}}}, \bibinfo {author} {\bibnamefont {{Majerotto,
  E.}}}, \bibinfo {author} {\bibnamefont {{Markovic, K.}}}, \bibinfo {author}
  {\bibnamefont {{Martinelli, M.}}}, \bibinfo {author} {\bibnamefont
  {{Pettorino, V.}}}, \bibinfo {author} {\bibnamefont {{Pourtsidou, A.}}},
  \bibinfo {author} {\bibnamefont {{Sakr, Z.}}}, \bibinfo {author}
  {\bibnamefont {{S\'anchez, A. G.}}}, \bibinfo {author} {\bibnamefont
  {{Sapone, D.}}}, \bibinfo {author} {\bibnamefont {{Tutusaus, I.}}},  \emph
  {et~al.},\ }\href {\doibase 10.1051/0004-6361/202038071} {\bibfield
  {journal} {\bibinfo  {journal} {A\&A}\ }\textbf {\bibinfo {volume} {642}},\
  \bibinfo {pages} {A191} (\bibinfo {year} {2020})}\BibitemShut {NoStop}%
\bibitem [{\citenamefont {Coulton}\ \emph
  {et~al.}(2020{\natexlab{a}})\citenamefont {Coulton}, \citenamefont {Liu},
  \citenamefont {McCarthy},\ and\ \citenamefont
  {Osato}}]{10.1093/mnras/staa1098}%
  \BibitemOpen
  \bibfield  {author} {\bibinfo {author} {\bibfnamefont {W.~R.}\ \bibnamefont
  {Coulton}}, \bibinfo {author} {\bibfnamefont {J.}~\bibnamefont {Liu}},
  \bibinfo {author} {\bibfnamefont {I.~G.}\ \bibnamefont {McCarthy}}, \ and\
  \bibinfo {author} {\bibfnamefont {K.}~\bibnamefont {Osato}},\ }\href
  {\doibase 10.1093/mnras/staa1098} {\bibfield  {journal} {\bibinfo  {journal}
  {Monthly Notices of the Royal Astronomical Society}\ }\textbf {\bibinfo
  {volume} {495}},\ \bibinfo {pages} {2531} (\bibinfo {year}
  {2020}{\natexlab{a}})}\BibitemShut {NoStop}%
\bibitem [{\citenamefont {{Starck}}\ \emph {et~al.}(2007)\citenamefont
  {{Starck}}, \citenamefont {{Fadili}},\ and\ \citenamefont
  {{Murtagh}}}]{4060954}%
  \BibitemOpen
  \bibfield  {author} {\bibinfo {author} {\bibfnamefont {J.}~\bibnamefont
  {{Starck}}}, \bibinfo {author} {\bibfnamefont {J.}~\bibnamefont {{Fadili}}},
  \ and\ \bibinfo {author} {\bibfnamefont {F.}~\bibnamefont {{Murtagh}}},\
  }\href {\doibase 10.1109/TIP.2006.887733} {\bibfield  {journal} {\bibinfo
  {journal} {IEEE Transactions on Image Processing}\ }\textbf {\bibinfo
  {volume} {16}},\ \bibinfo {pages} {297} (\bibinfo {year} {2007})}\BibitemShut
  {NoStop}%
\bibitem [{\citenamefont {Starck}\ \emph {et~al.}(1998)\citenamefont {Starck},
  \citenamefont {Murtagh},\ and\ \citenamefont
  {Bijaoui}}]{starck_murtagh_bijaoui_1998}%
  \BibitemOpen
  \bibfield  {author} {\bibinfo {author} {\bibfnamefont {J.-L.}\ \bibnamefont
  {Starck}}, \bibinfo {author} {\bibfnamefont {F.~D.}\ \bibnamefont {Murtagh}},
  \ and\ \bibinfo {author} {\bibfnamefont {A.}~\bibnamefont {Bijaoui}},\
  }\enquote {\bibinfo {title} {The wavelet transform},}\ in\ \href {\doibase
  10.1017/CBO9780511564352.002} {\emph {\bibinfo {booktitle} {Image Processing
  and Data Analysis: The Multiscale Approach}}}\ (\bibinfo  {publisher}
  {Cambridge University Press},\ \bibinfo {year} {1998})\ p.\ \bibinfo {pages}
  {1–45}\BibitemShut {NoStop}%
\bibitem [{\citenamefont {{Peel, Austin}}\ \emph {et~al.}(2018)\citenamefont
  {{Peel, Austin}}, \citenamefont {{Pettorino, Valeria}}, \citenamefont
  {{Giocoli, Carlo}}, \citenamefont {{Starck, Jean-Luc}},\ and\ \citenamefont
  {{Baldi, Marco}}}]{Peel2018}%
  \BibitemOpen
  \bibfield  {author} {\bibinfo {author} {\bibnamefont {{Peel, Austin}}},
  \bibinfo {author} {\bibnamefont {{Pettorino, Valeria}}}, \bibinfo {author}
  {\bibnamefont {{Giocoli, Carlo}}}, \bibinfo {author} {\bibnamefont {{Starck,
  Jean-Luc}}}, \ and\ \bibinfo {author} {\bibnamefont {{Baldi, Marco}}},\
  }\href {\doibase 10.1051/0004-6361/201833481} {\bibfield  {journal} {\bibinfo
   {journal} {A\&A}\ }\textbf {\bibinfo {volume} {619}},\ \bibinfo {pages}
  {A38} (\bibinfo {year} {2018})}\BibitemShut {NoStop}%
\bibitem [{job()}]{joblib}%
  \BibitemOpen
  \href@noop {} {}\bibinfo {note}
  {\url{https://joblib.readthedocs.io/en/latest/}}\BibitemShut {NoStop}%
\bibitem [{\citenamefont {Jee}\ \emph {et~al.}(2013)\citenamefont {Jee},
  \citenamefont {Tyson}, \citenamefont {Schneider}, \citenamefont {Wittman},
  \citenamefont {Schmidt},\ and\ \citenamefont {Hilbert}}]{Jee_2013}%
  \BibitemOpen
  \bibfield  {author} {\bibinfo {author} {\bibfnamefont {M.~J.}\ \bibnamefont
  {Jee}}, \bibinfo {author} {\bibfnamefont {J.~A.}\ \bibnamefont {Tyson}},
  \bibinfo {author} {\bibfnamefont {M.~D.}\ \bibnamefont {Schneider}}, \bibinfo
  {author} {\bibfnamefont {D.}~\bibnamefont {Wittman}}, \bibinfo {author}
  {\bibfnamefont {S.}~\bibnamefont {Schmidt}}, \ and\ \bibinfo {author}
  {\bibfnamefont {S.}~\bibnamefont {Hilbert}},\ }\href {\doibase
  10.1088/0004-637x/765/1/74} {\bibfield  {journal} {\bibinfo  {journal} {The
  Astrophysical Journal}\ }\textbf {\bibinfo {volume} {765}},\ \bibinfo {pages}
  {74} (\bibinfo {year} {2013})}\BibitemShut {NoStop}%
\bibitem [{\citenamefont {Abbott}\ \emph {et~al.}(2016)\citenamefont {Abbott},
  \citenamefont {Abdalla}, \citenamefont {Allam}, \citenamefont {Amara},
  \citenamefont {Annis}, \citenamefont {Armstrong}, \citenamefont {Bacon},
  \citenamefont {Banerji}, \citenamefont {Bauer}, \citenamefont {Baxter},
  \citenamefont {Becker}, \citenamefont {Benoit-L\'evy}, \citenamefont
  {Bernstein}, \citenamefont {Bernstein} \emph {et~al.}}]{PhysRevD.94.022001}%
  \BibitemOpen
  \bibfield  {author} {\bibinfo {author} {\bibfnamefont {T.}~\bibnamefont
  {Abbott}}, \bibinfo {author} {\bibfnamefont {F.~B.}\ \bibnamefont {Abdalla}},
  \bibinfo {author} {\bibfnamefont {S.}~\bibnamefont {Allam}}, \bibinfo
  {author} {\bibfnamefont {A.}~\bibnamefont {Amara}}, \bibinfo {author}
  {\bibfnamefont {J.}~\bibnamefont {Annis}}, \bibinfo {author} {\bibfnamefont
  {R.}~\bibnamefont {Armstrong}}, \bibinfo {author} {\bibfnamefont
  {D.}~\bibnamefont {Bacon}}, \bibinfo {author} {\bibfnamefont
  {M.}~\bibnamefont {Banerji}}, \bibinfo {author} {\bibfnamefont {A.~H.}\
  \bibnamefont {Bauer}}, \bibinfo {author} {\bibfnamefont {E.}~\bibnamefont
  {Baxter}}, \bibinfo {author} {\bibfnamefont {M.~R.}\ \bibnamefont {Becker}},
  \bibinfo {author} {\bibfnamefont {A.}~\bibnamefont {Benoit-L\'evy}}, \bibinfo
  {author} {\bibfnamefont {R.~A.}\ \bibnamefont {Bernstein}}, \bibinfo {author}
  {\bibnamefont {Bernstein}},  \emph {et~al.} (\bibinfo {collaboration} {The
  Dark Energy Survey Collaboration}),\ }\href {\doibase
  10.1103/PhysRevD.94.022001} {\bibfield  {journal} {\bibinfo  {journal} {Phys.
  Rev. D}\ }\textbf {\bibinfo {volume} {94}},\ \bibinfo {pages} {022001}
  (\bibinfo {year} {2016})}\BibitemShut {NoStop}%
\bibitem [{\citenamefont {Hildebrandt}\ \emph {et~al.}(2016)\citenamefont
  {Hildebrandt}, \citenamefont {Viola}, \citenamefont {Heymans}, \citenamefont
  {Joudaki}, \citenamefont {Kuijken}, \citenamefont {Blake}, \citenamefont
  {Erben}, \citenamefont {Joachimi}, \citenamefont {Klaes}, \citenamefont
  {Miller}, \citenamefont {Morrison}, \citenamefont {Nakajima}, \citenamefont
  {Verdoes~Kleijn}, \citenamefont {Amon}, \citenamefont {Choi} \emph
  {et~al.}}]{10.1093/mnras/stw2805}%
  \BibitemOpen
  \bibfield  {author} {\bibinfo {author} {\bibfnamefont {H.}~\bibnamefont
  {Hildebrandt}}, \bibinfo {author} {\bibfnamefont {M.}~\bibnamefont {Viola}},
  \bibinfo {author} {\bibfnamefont {C.}~\bibnamefont {Heymans}}, \bibinfo
  {author} {\bibfnamefont {S.}~\bibnamefont {Joudaki}}, \bibinfo {author}
  {\bibfnamefont {K.}~\bibnamefont {Kuijken}}, \bibinfo {author} {\bibfnamefont
  {C.}~\bibnamefont {Blake}}, \bibinfo {author} {\bibfnamefont
  {T.}~\bibnamefont {Erben}}, \bibinfo {author} {\bibfnamefont
  {B.}~\bibnamefont {Joachimi}}, \bibinfo {author} {\bibfnamefont
  {D.}~\bibnamefont {Klaes}}, \bibinfo {author} {\bibfnamefont
  {L.}~\bibnamefont {Miller}}, \bibinfo {author} {\bibfnamefont {C.~B.}\
  \bibnamefont {Morrison}}, \bibinfo {author} {\bibfnamefont {R.}~\bibnamefont
  {Nakajima}}, \bibinfo {author} {\bibfnamefont {G.}~\bibnamefont
  {Verdoes~Kleijn}}, \bibinfo {author} {\bibfnamefont {A.}~\bibnamefont
  {Amon}}, \bibinfo {author} {\bibnamefont {Choi}},  \emph {et~al.},\ }\href
  {\doibase 10.1093/mnras/stw2805} {\bibfield  {journal} {\bibinfo  {journal}
  {Monthly Notices of the Royal Astronomical Society}\ }\textbf {\bibinfo
  {volume} {465}},\ \bibinfo {pages} {1454} (\bibinfo {year}
  {2016})}\BibitemShut {NoStop}%
\bibitem [{len()}]{lenspack}%
  \BibitemOpen
  \href@noop {} {}\bibinfo {note}
  {\url{https://github.com/CosmoStat/lenspack}}\BibitemShut {NoStop}%
\bibitem [{\citenamefont {Starck}\ and\ \citenamefont
  {Murtagh}(1998)}]{Starck_1998}%
  \BibitemOpen
  \bibfield  {author} {\bibinfo {author} {\bibfnamefont {J.-L.}\ \bibnamefont
  {Starck}}\ and\ \bibinfo {author} {\bibfnamefont {F.}~\bibnamefont
  {Murtagh}},\ }\href {\doibase 10.1086/316124} {\bibfield  {journal} {\bibinfo
   {journal} {Publications of the Astronomical Society of the Pacific}\
  }\textbf {\bibinfo {volume} {110}},\ \bibinfo {pages} {193} (\bibinfo {year}
  {1998})}\BibitemShut {NoStop}%
\bibitem [{\citenamefont {Liu}\ and\ \citenamefont
  {Haiman}(2016)}]{PhysRevD.94.043533}%
  \BibitemOpen
  \bibfield  {author} {\bibinfo {author} {\bibfnamefont {J.}~\bibnamefont
  {Liu}}\ and\ \bibinfo {author} {\bibfnamefont {Z.}~\bibnamefont {Haiman}},\
  }\href {\doibase 10.1103/PhysRevD.94.043533} {\bibfield  {journal} {\bibinfo
  {journal} {Phys. Rev. D}\ }\textbf {\bibinfo {volume} {94}},\ \bibinfo
  {pages} {043533} (\bibinfo {year} {2016})}\BibitemShut {NoStop}%
\bibitem [{\citenamefont {Rasmussen}\ and\ \citenamefont
  {Williams}(2005)}]{Rasmussen:2005:GPM:1162254}%
  \BibitemOpen
  \bibfield  {author} {\bibinfo {author} {\bibfnamefont {C.~E.}\ \bibnamefont
  {Rasmussen}}\ and\ \bibinfo {author} {\bibfnamefont {C.~K.~I.}\ \bibnamefont
  {Williams}},\ }\href@noop {} {\emph {\bibinfo {title} {Gaussian Processes for
  Machine Learning (Adaptive Computation and Machine Learning)}}}\ (\bibinfo
  {publisher} {The MIT Press},\ \bibinfo {year} {2005})\BibitemShut {NoStop}%
\bibitem [{\citenamefont {{Carron, J.}}(2013)}]{2013A&A...551A..88C}%
  \BibitemOpen
  \bibfield  {author} {\bibinfo {author} {\bibnamefont {{Carron, J.}}},\ }\href
  {\doibase 10.1051/0004-6361/201220538} {\bibfield  {journal} {\bibinfo
  {journal} {A\&A}\ }\textbf {\bibinfo {volume} {551}},\ \bibinfo {pages} {A88}
  (\bibinfo {year} {2013})}\BibitemShut {NoStop}%
\bibitem [{\citenamefont {{Hartlap, J.}}\ \emph {et~al.}(2007)\citenamefont
  {{Hartlap, J.}}, \citenamefont {{Simon, P.}},\ and\ \citenamefont
  {{Schneider, P.}}}]{2007A&A...464..399H}%
  \BibitemOpen
  \bibfield  {author} {\bibinfo {author} {\bibnamefont {{Hartlap, J.}}},
  \bibinfo {author} {\bibnamefont {{Simon, P.}}}, \ and\ \bibinfo {author}
  {\bibnamefont {{Schneider, P.}}},\ }\href {\doibase
  10.1051/0004-6361:20066170} {\bibfield  {journal} {\bibinfo  {journal}
  {A\&A}\ }\textbf {\bibinfo {volume} {464}},\ \bibinfo {pages} {399} (\bibinfo
  {year} {2007})}\BibitemShut {NoStop}%
\bibitem [{\citenamefont {Sellentin}\ and\ \citenamefont
  {Heavens}(2015)}]{10.1093/mnrasl/slv190}%
  \BibitemOpen
  \bibfield  {author} {\bibinfo {author} {\bibfnamefont {E.}~\bibnamefont
  {Sellentin}}\ and\ \bibinfo {author} {\bibfnamefont {A.~F.}\ \bibnamefont
  {Heavens}},\ }\href {\doibase 10.1093/mnrasl/slv190} {\bibfield  {journal}
  {\bibinfo  {journal} {Monthly Notices of the Royal Astronomical Society:
  Letters}\ }\textbf {\bibinfo {volume} {456}},\ \bibinfo {pages} {L132}
  (\bibinfo {year} {2015})}\BibitemShut {NoStop}%
\bibitem [{\citenamefont {Sellentin}\ and\ \citenamefont
  {Heavens}(2016)}]{10.1093/mnras/stw2697}%
  \BibitemOpen
  \bibfield  {author} {\bibinfo {author} {\bibfnamefont {E.}~\bibnamefont
  {Sellentin}}\ and\ \bibinfo {author} {\bibfnamefont {A.~F.}\ \bibnamefont
  {Heavens}},\ }\href {\doibase 10.1093/mnras/stw2697} {\bibfield  {journal}
  {\bibinfo  {journal} {Monthly Notices of the Royal Astronomical Society}\
  }\textbf {\bibinfo {volume} {464}},\ \bibinfo {pages} {4658} (\bibinfo {year}
  {2016})}\BibitemShut {NoStop}%
\bibitem [{\citenamefont {Casas}\ \emph {et~al.}(2017)\citenamefont {Casas},
  \citenamefont {Kunz}, \citenamefont {Martinelli},\ and\ \citenamefont
  {Pettorino}}]{CASAS201773}%
  \BibitemOpen
  \bibfield  {author} {\bibinfo {author} {\bibfnamefont {S.}~\bibnamefont
  {Casas}}, \bibinfo {author} {\bibfnamefont {M.}~\bibnamefont {Kunz}},
  \bibinfo {author} {\bibfnamefont {M.}~\bibnamefont {Martinelli}}, \ and\
  \bibinfo {author} {\bibfnamefont {V.}~\bibnamefont {Pettorino}},\ }\href
  {\doibase https://doi.org/10.1016/j.dark.2017.09.009} {\bibfield  {journal}
  {\bibinfo  {journal} {Physics of the Dark Universe}\ }\textbf {\bibinfo
  {volume} {18}},\ \bibinfo {pages} {73 } (\bibinfo {year} {2017})}\BibitemShut
  {NoStop}%
\bibitem [{\citenamefont {{Foreman-Mackey}}\ \emph {et~al.}(2013)\citenamefont
  {{Foreman-Mackey}}, \citenamefont {{Hogg}}, \citenamefont {{Lang}},\ and\
  \citenamefont {{Goodman}}}]{2013PASP..125..306F}%
  \BibitemOpen
  \bibfield  {author} {\bibinfo {author} {\bibfnamefont {D.}~\bibnamefont
  {{Foreman-Mackey}}}, \bibinfo {author} {\bibfnamefont {D.~W.}\ \bibnamefont
  {{Hogg}}}, \bibinfo {author} {\bibfnamefont {D.}~\bibnamefont {{Lang}}}, \
  and\ \bibinfo {author} {\bibfnamefont {J.}~\bibnamefont {{Goodman}}},\ }\href
  {\doibase 10.1086/670067} {\bibfield  {journal} {\bibinfo  {journal} {\pasp}\
  }\textbf {\bibinfo {volume} {125}},\ \bibinfo {pages} {306} (\bibinfo {year}
  {2013})},\ \Eprint {http://arxiv.org/abs/1202.3665} {arXiv:1202.3665
  [astro-ph.IM]} \BibitemShut {NoStop}%
\bibitem [{\citenamefont {Gelman}\ and\ \citenamefont
  {Rubin}(1992)}]{gelman1992}%
  \BibitemOpen
  \bibfield  {author} {\bibinfo {author} {\bibfnamefont {A.}~\bibnamefont
  {Gelman}}\ and\ \bibinfo {author} {\bibfnamefont {D.~B.}\ \bibnamefont
  {Rubin}},\ }\href {\doibase 10.1214/ss/1177011136} {\bibfield  {journal}
  {\bibinfo  {journal} {Statist. Sci.}\ }\textbf {\bibinfo {volume} {7}},\
  \bibinfo {pages} {457} (\bibinfo {year} {1992})}\BibitemShut {NoStop}%
\bibitem [{\citenamefont {Hinton}(2016)}]{Hinton2016}%
  \BibitemOpen
  \bibfield  {author} {\bibinfo {author} {\bibfnamefont {S.}~\bibnamefont
  {Hinton}},\ }\href {\doibase 10.21105/joss.00045} {\bibfield  {journal}
  {\bibinfo  {journal} {Journal of Open Source Software}\ }\textbf {\bibinfo
  {volume} {1}},\ \bibinfo {pages} {45} (\bibinfo {year} {2016})}\BibitemShut
  {NoStop}%
\bibitem [{\citenamefont {Leonard}\ \emph {et~al.}(2012)\citenamefont
  {Leonard}, \citenamefont {Pires},\ and\ \citenamefont
  {Starck}}]{2012MNRAS.423.3405L}%
  \BibitemOpen
  \bibfield  {author} {\bibinfo {author} {\bibfnamefont {A.}~\bibnamefont
  {Leonard}}, \bibinfo {author} {\bibfnamefont {S.}~\bibnamefont {Pires}}, \
  and\ \bibinfo {author} {\bibfnamefont {J.-L.}\ \bibnamefont {Starck}},\
  }\href {\doibase 10.1111/j.1365-2966.2012.21133.x} {\bibfield  {journal}
  {\bibinfo  {journal} {Monthly Notices of the Royal Astronomical Society}\
  }\textbf {\bibinfo {volume} {423}},\ \bibinfo {pages} {3405} (\bibinfo {year}
  {2012})}\BibitemShut {NoStop}%
\bibitem [{\citenamefont {Fong}\ \emph {et~al.}(2019)\citenamefont {Fong},
  \citenamefont {Choi}, \citenamefont {Catlett}, \citenamefont {Lee},
  \citenamefont {Peel}, \citenamefont {Bowyer}, \citenamefont {King},\ and\
  \citenamefont {McCarthy}}]{2019MNRAS.488.3340F}%
  \BibitemOpen
  \bibfield  {author} {\bibinfo {author} {\bibfnamefont {M.}~\bibnamefont
  {Fong}}, \bibinfo {author} {\bibfnamefont {M.}~\bibnamefont {Choi}}, \bibinfo
  {author} {\bibfnamefont {V.}~\bibnamefont {Catlett}}, \bibinfo {author}
  {\bibfnamefont {B.}~\bibnamefont {Lee}}, \bibinfo {author} {\bibfnamefont
  {A.}~\bibnamefont {Peel}}, \bibinfo {author} {\bibfnamefont {R.}~\bibnamefont
  {Bowyer}}, \bibinfo {author} {\bibfnamefont {L.~J.}\ \bibnamefont {King}}, \
  and\ \bibinfo {author} {\bibfnamefont {I.~G.}\ \bibnamefont {McCarthy}},\
  }\href {\doibase 10.1093/mnras/stz1882} {\bibfield  {journal} {\bibinfo
  {journal} {Monthly Notices of the Royal Astronomical Society}\ }\textbf
  {\bibinfo {volume} {488}},\ \bibinfo {pages} {3340} (\bibinfo {year}
  {2019})}\BibitemShut {NoStop}%
\bibitem [{\citenamefont {Coulton}\ \emph
  {et~al.}(2020{\natexlab{b}})\citenamefont {Coulton}, \citenamefont {Liu},
  \citenamefont {McCarthy},\ and\ \citenamefont {Osato}}]{2020MNRAS.495.2531C}%
  \BibitemOpen
  \bibfield  {author} {\bibinfo {author} {\bibfnamefont {W.~R.}\ \bibnamefont
  {Coulton}}, \bibinfo {author} {\bibfnamefont {J.}~\bibnamefont {Liu}},
  \bibinfo {author} {\bibfnamefont {I.~G.}\ \bibnamefont {McCarthy}}, \ and\
  \bibinfo {author} {\bibfnamefont {K.}~\bibnamefont {Osato}},\ }\href
  {\doibase 10.1093/mnras/staa1098} {\bibfield  {journal} {\bibinfo  {journal}
  {Monthly Notices of the Royal Astronomical Society}\ }\textbf {\bibinfo
  {volume} {495}},\ \bibinfo {pages} {2531} (\bibinfo {year}
  {2020}{\natexlab{b}})}\BibitemShut {NoStop}%
\end{thebibliography}%

\end{document}